\documentclass[twocolumn,aps,prb,showpacs,raggedbottom,nobalancelastpage,amssymb,superscriptaddress,citeautoscript]{revtex4-1}

\usepackage{epsfig,amssymb,amsmath,latexsym}
\usepackage{graphics}
\usepackage{subfigure}
\usepackage{wrapfig}
\usepackage{float}
\usepackage{dsfont}
\usepackage[usenames]{color}
\usepackage{soul}
\setulcolor{red}
\setstcolor{red}
\sethlcolor{yellow}

\usepackage{varioref}
\usepackage{subfigure} 
\usepackage{amssymb, amsmath, amsfonts}
\usepackage{pst-plot, pstricks}
\usepackage{setspace}
\usepackage{calc}
\usepackage{array}
\usepackage{layout}
\usepackage{url}
\usepackage{rotating} 
\usepackage{fancybox,amssymb,amsmath}
\usepackage[usenames]{color}
\usepackage{graphicx} 
\usepackage{wasysym}
\usepackage{bbm}
\usepackage{ragged2e} 
\usepackage{hyperref}

\usepackage[title]{appendix}

\begin{document}
\title{Langevin dynamics of a heavy particle and orthogonality effects}
\author{Mark Thomas}
\affiliation{\mbox{Dahlem Center for Complex Quantum Systems and Fachbereich Physik, Freie Universit\"at Berlin, 14195 Berlin, Germany}}
\author{Torsten Karzig}
\affiliation{Department of Physics, California Institute of Technology, Pasadena, California 91125, USA}
\author{Silvia Viola Kusminskiy}
\affiliation{\mbox{Dahlem Center for Complex Quantum Systems and Fachbereich Physik, Freie Universit\"at Berlin, 14195 Berlin, Germany}}

\date{\today}
\begin{abstract}
The dynamics of a classical heavy particle moving in a quantum environment is
determined by a Langevin equation which encapsulates the effect of the
environment-induced reaction forces on the particle. For an open quantum system
these include a Born-Oppenheimer force, a dissipative force and a stochastic force
due to shot and thermal noise. Recently it was shown that these forces can be
expressed in terms of the scattering matrix of the system by considering the
classical heavy particle as a time-dependent scattering center, allowing to demonstrate interesting features of these forces when the system is driven out of equilibrium. At the same time, it
is well known that small changes in a scattering potential can have a profound
impact on a fermionic system due to the Anderson orthogonality catastrophe. In this work, by calculating the Loschmidt echo, we relate Anderson orthogonality effects with the mesoscopic reaction forces for an environment that can be taken out of equilibrium. In particular we show how the decay of the Loschmidt echo is characterized by fluctuations and dissipation in the system and discuss different quench protocols.  
\end{abstract}

\maketitle

%%%%%%%%%%%%%%%%%%%%%%%%%%%%%%%%%%%%%%%%%%%%%%%%%%%%%%%%%%%%%%%%%%

\section{Introduction}

Understanding the effect of fluctuations and dissipation in non-equilibrium settings is essential for an ultimate control of quantum systems. Dissipation is on one hand unavoidable in realistic systems, and known to play an important role in their dynamics --- a paradigm is the exponential supression of quantum tunneling out of a metastable state as modeled by Caldeira and Leggett~\cite{Caldeira:1981, Caldeira:1983} --- while non-equilibrium can provide new levels of tunability. This is a topic of renewed interest in view of current experiments which explore the possibility of quantum information processing, by embedding a qubit degree of freedom in a mesoscopic system~\cite{DevoretScience13,AwschalomScience13}. The coupling of the qubit to an environment causes decoherence and consequently loss of information, which is closely related to the fluctuations and dissipation in the system.~\cite{Weiss:1999}

In this context, the quantum Loschmidt echo, also known as {\it fidelity}, is a useful quantity that indicates the sensitivity of the system to small perturbations.~\cite{Peres:1984,Gorin:2006_PhysRep} In its generalization to many-body systems,~\cite{Lesovik:2006} the Loschmidt echo corresponds to the off-diagonal element (norm-squared) of the reduced density matrix for the qubit degree of freedom, and its decay in time characterizes the environment-induced decoherence.~\cite{Karkuszewski:2002} For a fermionic environment, this decay is directly related to the Anderson 
orthogonality catastrophe, which describes the response of the fermionic system to a sudden perturbation.~\cite{Goold:2011} In his seminal work,~\cite{Anderson:1967} Anderson showed that the many-body ground state of a fermionic system is, in the thermodynamic limit, orthogonal to that of the same system in which a local scattering potential is introduced. More precisely, the overlap of the two states decays as a powerlaw with the system size, with an orthogonality exponent characterized by the scattering phase shift produced by the scattering potential. The orthogonality catastrophe plays an essential role in describing the so-called ``impurity problems'' in which a local degree of freedom interacts with a fermionic environment.~\cite{Mahan:1967, Dominicis:1969, Rivier:1971, Ng:1995, Ng:1996, Levitov:2004, Levitov:2005}

A class of impurity problems is that of a ``heavy particle'' embedded in a quantum environment. The impurity in this case is heavy compared to those particles comprising the environment, and can be treated as a classical degree of freedom with semiclassical dynamics dictated by the back-action of the environment. This dynamics can be described in terms of a Langevin equation, which is a stochastic equation of motion that describes at an effective, macroscopic level, the effects of dissipation and fluctuations induced by the environment on the heavy particle. An interesting question is how the dynamics of the heavy particle and orthogonality effects of its environment are related. For a quantum environment in equilibrium, the Anderson orthogonality catastrophe exponent was conjectured by Sols and Guinea~\cite{Sols:1987} to be proportional to the dissipation coefficient a heavy particle experiences when moving in a metallic environment. This relation was later proved to be valid, in the small-distance limit,~\footnote{The 
small-distance 
limit 
corresponds to small variations in the classical coordinates.} for a heavy particle moving in a quantum environment at zero temperature.~\cite{Schoenhammer:1991} For non-equilibrium fermionic systems this problem has been studied in the context of concrete models~\cite{Ng:1995,Ng:1996,Braunecker:2003,Levitov:2005,Segal:2007}.

Motivated by these findings, in this work we calculate the Loschmidt echo for small changes of a scattering potential, in a fermionic open system which is taken out of equilibrium by imposing a voltage bias. With the aim of exploring the relation between the orthogonality exponent and the dissipation coefficient in this case, we express the decay in time of the Loschmidt echo in terms of the coefficients of the corresponding Langevin equation - in particular in terms of the dissipation and noise coefficients. To this effect we make use of the recent developed formalism that describes the effective forces in the Langevin equation in terms of scattering theory.~\cite{Bode:2011,Thomas:2012} Our results apply generally to systems for which changes in the scattering potential can be treated pertubatively. 

The manuscript is organized as follows. We start in section Sec.~\ref{sec:LangevinAndNoise} by presenting the Langevin equation in terms of current-induced forces, and the associated force-force noise correlator. In Sec.~\ref{sec:LE} we perform a perturbative expansion of the Loschmidt echo and show that it can be expressed in terms of the noise correlator, and discuss different quench protocols. In Sec.~\ref{sec:Equilibrium} we make use of the results of Secs.~\ref{sec:LangevinAndNoise}  and~\ref{sec:LE} to show that, in equilibrium and for zero temperature, the decay of the Loschmidt echo is a powerlaw with an exponent dictated by the dissipation coefficient a heavy particle experiences in the fermionic environment, in agreement with the known orthogonality results. Finite temperatures  however render the decay exponential. In Sec.~\ref{sec:nonequilibrium} we turn to 
the non-equilibrium case for which we calulate the decay of the Loschmidt echo within linear response in the applied bias. In this case we show that the decay of the Loschmidt echo cannot be expressed solely in terms of the dissipation coefficient, providing a general expression for the decay in terms of the macroscopic Langevin parameters. We then discuss different time scales for which the results can be cast in a simple form. In Sec.~\ref{sec:2-level_model} we apply our results to a simple example and check the limits of validity of our approximations, while we list our main conclusions in Sec.~\ref{sec:conclusion}. Quite a few calculations in this work are rather lenghty. To improve readability, and at the same time to make this paper self contained, we have included some details of these calculations in the Appendices.

\section{Langevin Equation and Noise Correlator}\label{sec:LangevinAndNoise}
\begin{figure}[t]
\begin{centering}
\includegraphics[width=0.45\textwidth]{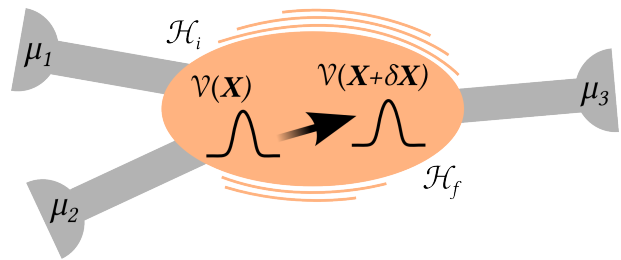}
\end{centering}
\caption{In the scattering region electrons couple ${\bf X}$ via the scattering potential $\mathcal{V}_{{\bf X}}$. When the scatterer moves by $\delta {\bf X}$, the scattering potential changes accordingly, i.e. $\mathcal{V}_{{\bf X}+ \delta{\bf X}}$. This gives rise to the two different Hamiltonians $\mathcal{H}_\mathrm{i} = \mathcal{H}_0 + \mathcal{V}_{{\bf X}}$ and $\mathcal{H}_\mathrm{f} = \mathcal{H}_0  + \mathcal{V}_{{\bf X}+ \delta{\bf X}}$. The scattering region is finite and the dwell time $\tau_D$ gives the timescale the electrons spend within.}
\label{fig:mov_scatt}
\end{figure}

In this section we briefly review the elements of the Langevin equation that governs the stochastic dynamics of a heavy particle in an open quantum environment. Throughout this paper we will consider a fermionic quantum environment that can be taken out of equilibrium by a difference of chemical potential in the leads as illustrated in Fig.~\ref{fig:mov_scatt}. In a concrete example, the heavy particle can represent the classical vibrational degrees of freedom of a molecule or suspended carbon nanotube connected to conducting leads. The heavy particle is represented through classical degrees of freedom which are coupled to the quantum environment, and disturb it as they evolve in time. The back-action of this disturbance onto the heavy particle gives rise to reaction forces~\cite{Berry:1993}, also called current-induced forces in a quantum transport setup. In the adiabatic limit, for which the dynamics of the heavy particle is much slower that that of the quantum environment, this effect is well described 
semiclassically at the level of a Langevin 
equation 
obtained by tracing out the quantum environment. If we denote the degrees of freedom of the heavy particle by $\mathbf{X}(t)$, the Langevin equation reads (in what follows we ommit the time dependence for notational simplicity)
\begin{align}\label{eq:Langevin}
\dot{P}_\alpha -F^{\rm{cl}}_\alpha(\mathbf{X}) = F_\alpha(\mathbf{X})  - \sum_{\beta} \Gamma_{\alpha\beta}(\mathbf{X})  \dot{X}_{\beta} +\xi_\alpha(\mathbf{X}) \,.
\end{align}
On the left hand side, $P_\alpha$ denotes the canonical momentum of coordinate $X_\alpha$ $(\alpha=1,\ldots,N)$, and we have included the possibility of an external classical force $\mathbf{F}^{\rm{cl}}({\bf X})$ (troughout the text we will indicate matrices and vectors in the space spanned by the $X_\alpha$ with bold letters). The right hand side of  Eq.~\eqref{eq:Langevin} contains the forces due to the quantum environment. $\mathbf{F}({\bf X})$ is the usual Born-Oppenheimer force, while the symmetric and antisymmetric parts of the tensor $\boldsymbol{\Gamma}({\bf X})$ represent a dissipative and Lorentz-like (and therefore non-dissipative) force, respectively. Fluctuations due to shot and thermal noise are taken into account by the stochastic Langevin force~$\boldsymbol{\xi}({\bf X})$.  

When the fermionic system is taken out of equilibrium, the current-induced forces present qualitative differences with respect to the equilibrium situation~\cite{Todorov:2009,Hedegard:2010,Bode:2011, Bode:2012}. The Born-Oppenheimer force $\mathbf{F}$ is non-conservative in this case, and therefore provides a way of exchanging energy between the classical field and the quantum environment which is non-dissipative. The tensor $\boldsymbol{\Gamma}$ constitutes the first order correction in an adiabatic expansion to the Born-Oppenheimer force. It can be split into symmetric and antisymmetric components. The antisymmetric component is a Lorentz-like term, which can be interpreted as an effective magnetic field acting on the space spanned by $\mathbf{X}$ --- this term is not relevant for the Loschmidt echo (which involves only symmetric components as we will se below) and hence will be not dealt further with within this work.  We denote the symmetric dissipative term of $\boldsymbol{\Gamma}$ by $\boldsymbol{\gamma}$. It is convenient to express the latter as $\boldsymbol{\gamma} = \boldsymbol{\gamma}^{\mathrm{\rm{eq}}} + \boldsymbol{\gamma}^{\mathrm{\rm{neq}}}$, where 
$\boldsymbol{\gamma}^{\mathrm{\rm{neq}}}$ 
represents a pure non-equilibrium contribution while $\boldsymbol{\gamma}^{\mathrm{\rm{eq}}}$ is a straightforward generalization of the equilibrium contribution evaluated in an non-equilibrium environment. Explicit expressions for these quantities are given in App.~\ref{app:scattering1}. According to our definition, $\boldsymbol{\gamma}^{\mathrm{\rm{eq}}}$ connects to the equilibrium results, but it also contains non-equilibrium terms for finite bias. The pure non-equilibrium term $\boldsymbol{\gamma}^{\mathrm{\rm{neq}}}$ can take negative values and, moreover, render the full dissipative term negative.  

Of particular importance for the following discussion is the stochastic component $\boldsymbol{\xi}$, characterized by the force-force noise correlator. We will see in the next section that the Loschmidt echo can be, perturbatively for small displacements $\delta{\bf X}$, written in terms of the noise correlator
\begin{align}\label{eq:ColorNoise}
 D_{\alpha\beta}(t,t') = \big\{\big\langle\hat{\xi}_{\alpha}(t)\hat{\xi}_{\beta}(t')\big\rangle\big\}_{\rm{\rm{s}}}\,,
\end{align}
where $\left\{M_{\alpha\beta}\right\}_\mathrm{\rm{s}}=\left(M_{\alpha\beta}+M_{\beta\alpha}\right)/2$ indicates the symmetric component of a generic matrix $\mathbf{M}$. To give a concrete expression for this noise correlator we consider a generic, albeit non-interacting many-body Hamiltonian which depends parametrically on time {\it via} the potential $\mathcal{V}_{\mathbf{X}}$,  $\mathcal{H}_{\mathbf{X}} = \mathcal{H}_0 + \mathcal{V}_{\mathbf{X}}$. This potential represents the coupling between the heavy particle and the fermionic environment. The current-induced force operator is given by
\begin{align}\label{eq:ForceOperator}
\boldsymbol{\mathcal{F}}(t) = - \nabla_{\mathbf{X}} \mathcal{H}_{\mathbf{X}}(t)\,.
\end{align} 
The Langevin Eq.~\eqref{eq:Langevin} is obtained by calculating the quantum-statistical average $\langle \boldsymbol{\mathcal{F}}(t) \rangle$ within an adiabatic expansion to linear order in the velocity $\dot{\mathbf{X}}$, together with the quantum and thermal fluctuations given by
\begin{align}\label{eq:LangevinForce}
\hat{\boldsymbol{\mathbf{\xi}}} (t) = \boldsymbol{\mathcal{F}}(t) - \langle \boldsymbol{\mathcal{F}}(t) \rangle\,.
\end{align} 
The coefficients of this expansion are instantaneous: the noise is assumed to be delta correlated $\mathbf{D}(t,t')\rightarrow \mathbf{D}(\mathbf{X})\,\delta(t-t')$, and there is no retardation kernel for the dissipative term $\boldsymbol{\gamma}(\mathbf{X})$. These are the zero-frequency limit, respectively, of the force-force correlator~\eqref{eq:ColorNoise} and force susceptibility
\begin{align}
\chi^{FF}_{\alpha\beta}(t,t')=  - i \, \theta(t-t')  \langle \left[ \mathcal{F}_\alpha(t) , \, \mathcal{F}_\beta(t') \right] \rangle 
\end{align}
where $\theta(t)$ is the usual step function. In equilibrium and assuming steady state, so the time dependence in the relevant quantities is through the time difference $(t-t')$, we have 
\begin{align}
 \gamma_{\alpha \beta}^{\rm{eq}}(\omega) = - \frac{\textrm{Im} \,\chi^{FF}_{\alpha\beta}(\omega) }{\omega} \,.
\end{align} 
where $\gamma_{\alpha \beta}^{\rm{eq}}(\omega)$ denotes the Fourier transform of the friction Kernel $\gamma_{\alpha \beta}^{\rm{eq}}(t-t')$.~\footnote{We note that we assume the initial Hamiltonian $\mathcal{H}_{\mathbf{X}} = \mathcal{H}_0 + \mathcal{V}_{\mathbf{X}}$ to be time independent. The Green functions depend therefore only on the difference of the time arguments. Throughout the paper we define the Fourier transform of a function $f(t)$ as $f(\omega) = \int_{-\infty}^{\infty} \textrm{d}t \, e^{i \omega  t} \, f(t) $ with its inverse $f(t) = \int_{-\infty}^{\infty} \frac{\textrm{d}t}{2\pi} \, e^{-i \omega  t} \, f(\omega) $.}
The correlator of the fluctuating Langevin force and the friction tensor are related in equilibrium via the finite frequency fluctuation-dissipation theorem~\cite{Kubo:1966}
\begin{align}\label{eq:fluct-diss}
 \mathbf{D}(\omega) =  \omega \, \coth \left( \frac{\omega}{2 \,T} \right)\boldsymbol{\gamma}^{\mathrm{\rm{eq}}}(\omega) \, 
\end{align}
(we take the Boltzmann constant $k_B=1$), where $\mathbf{D}(\omega)$ is the real part of the Fourier transform of the fluctuating force correlator in Eq.~(\ref{eq:ColorNoise}). In the limit $\omega \ll T$ Eq.~\eqref{eq:fluct-diss} reduces to the classical identity 
\begin{align}\label{eq:FD_CL}
\mathbf{D}_\mathrm{cl}= 2\,T \, \boldsymbol{\gamma}^{\mathrm{\rm{eq}}}\,, 
\end{align}
where $\boldsymbol{\gamma}^{\mathrm{\rm{eq}}}$ is evaluated at zero frequency, while for $\omega \gg T$
\begin{align}\label{eq:FD_Q}
\mathbf{D}_{\rm{q}}(\omega)=|\omega| \,\boldsymbol{\gamma}^{\mathrm{\rm{eq}}}(\omega) \,.
\end{align}

Out of equilibrium the fluctuation-dissipation relation Eq.~\eqref{eq:fluct-diss} does not hold. However within linear response in the applied bias $\Delta \mu$ (we consider two leads and without loss of generality $\Delta\mu>0$), we can write an expression relating the noise correlator and the dissipative matrix $\boldsymbol{\gamma}$ that generalizes Eq.~\eqref{eq:fluct-diss} in the limit of low frequencies (as compared to the characteristic energy scales of the quantum environment, this statement will be made more precise in the following sections). We state here the result which will be proven later in the text
\begin{align}\label{eq:D_omega_sym_tot}
 \begin{split}
  \mathbf{D}(\omega) &= \omega \, \coth \left(\frac{\omega}{2\, T} \right) \,  \left(\boldsymbol{\gamma}^{\mathrm{\rm{eq}}} - \frac{\mathbf{D}_{[0,\Delta\mu]}}{\Delta\mu} \right)  \\
  &+ \left[ \frac{\omega_+}{2}  \coth \left( \frac{\omega_+}{2\, T} \right) + \frac{\omega_-}{2} \coth \left( \frac{\omega_- }{2\, T} \right) \right] \frac{\mathbf{D}_{[0,\Delta\mu]}}{\Delta\mu}   \\
  &+ \frac{\omega}{2} \left[ \omega_+\, \coth \left( \frac{\omega_+ }{2\, T} \right) - \omega_-  \, \coth \left( \frac{\omega_- }{2\, T} \right) \right] \,\frac{\boldsymbol{\gamma}^{\mathrm{\rm{neq}}}}{\Delta\mu}\,.
 \end{split}
\end{align}
where we defined $\omega_\pm=\omega \pm \Delta \mu$, and all remaining quantities are evaluated at zero frequency and to linear order in the applied bias. In general the noise correlator depends on both temperature and bias, which at zero frequency we denote by $\mathbf{D}_{[T,\Delta\mu]}$. It is easy to see that for $\Delta\mu = 0$ Eq.~\eqref{eq:D_omega_sym_tot} reduces to the equilibrium identity~\eqref{eq:fluct-diss}, and in particular $\mathbf{D}_{[T,0]}=\mathbf{D}_{\rm{cl}}$. For zero temperature but finite bias accordingly we obtain
\begin{align}\label{eq:D_mu_w}
\mathbf{ D}(\omega) = \left\{ 
 \begin{array}{ll}
 \mathbf{D}_{[0,\Delta\mu]}+|\omega| \, \left(  \boldsymbol{\gamma}^{\mathrm{\rm{eq}}}   - \frac{\mathbf{D}_{[0,\Delta\mu]}}{\Delta \mu} \right)  &, \quad |\omega| \ll \Delta\mu \\
  |\omega| \boldsymbol{\gamma}^{\mathrm{\rm{eq}}}  &, \quad |\omega| \gg \Delta\mu \\
 \end{array}
 \right. \,.
\end{align}
The explicit expression for $\mathbf{D}_{[0,\Delta\mu]}$ is given later in the text in Eq.~\eqref{eq:D_neq}.~\footnote{Note that the quantities  $\mathbf{D}$ and $\mathbf{\gamma}$ depend on $\mathbf{X}$. In order to simplify the notation, we do not write this dependence explicitly. } One should note the similarities and subtle differences between the finite temperature expressions, Eqs.~\eqref{eq:FD_CL}, \eqref{eq:FD_Q}, and finite bias Eq.~\eqref{eq:D_mu_w} in the respective limits of high and low bias and temperatures when identifying $\Delta\mu\leftrightarrow 2 T$ . All these expressions are evaluated to first order in $\omega$. From this we note that the first correction to Eq.~\eqref{eq:FD_CL} is of order $\mathcal{O}(\omega^2)$. That the linear term in $\omega$ vanishes in equilibrium is easy to see by looking at the corresponding finite bias result and identifying $\mathbf{D}_{[0,\Delta\mu]}\Delta\mu\leftrightarrow \mathbf{D}_{[T,0]}/2 T=\boldsymbol{\gamma}^{\mathrm{\rm{eq}}}$ in the first line of Eq.~\eqref{eq:D_mu_w}.

In the next section we show that the Loschmidt echo, within a perturbative expansion in the change of the scattering potential, is directly related to the noise correlator of Eq.~(\ref{eq:ColorNoise}).

\section{Loschmidt echo: general results for small displacements}\label{sec:LE}
The Loschmidt echo in quantum systems is given by the (squared) overlap of eigenstates of an initial system which evolved in time with two different many-body Hamiltonians. Alternatively, it can be seen as measure of how close to a given initial state a system comes back to, when the evolution on the time reversed path is determined by a different Hamiltonian from the forward evolution one. This can be generalized for initial states which are a quantum statistical mixture.~\cite{Lesovik:2006} Denoting the Loschmidt echo by the function $\mathcal{L}(\tau)$, this is given by $\mathcal{L}(\tau)=|\mathcal{A}(\tau)|^2$, with 
\begin{align}\label{eq:FA_def}
 \mathcal{A}(\tau) = \langle  \, e^{ i \mathcal{H}_\mathrm{i} \tau}\,e^{-i \mathcal{H}_\mathrm{f} \tau} \, \rangle \,,
\end{align}
where  $\langle \ldots \rangle $ is the quantum statistical average characterized by the initial Hamiltonian $\mathcal{H}_\mathrm{i}$ ($\hbar = 1$) and $\mathcal{H}_\mathrm{f}$ denotes the perturbed Hamiltonian. The overlap $\mathcal{A}(\tau)$ is denominated fidelity amplitude. 

The relation between the Loschmidt echo and the orthogonality catastrophe is seen by treating Anderson's orthogonality as a dynamical process.~\cite{Dominicis:1969} In the problem of X-ray absorption spectrum of a 
metal, the creation of a deep hole produces a ``shake up'' of the Fermi sea that causes a  suppression of Mahan's powerlaw divergence at treshold frequency (known as the Fermi edge or X-ray singularity),~\cite{Mahan:1967} with an exponent that can be identified directly with Anderson's orthogonality exponent~\cite{Rivier:1971}. This is captured by the hole propagator which can be calculated by evaluating the overlap of the fermionic ground state evolved with a Hamiltonian including the core hole, with that of the ground state evolved without the hole. This is therefore nothing else than the Loschmidt echo where the two Hamiltonians $\mathcal{H}_\mathrm{i}$, $\mathcal{H}_\mathrm{f}$ correspond to considering the system with or without the potential of the core hole. Beyond the original problem of the Fermi edge singularity in metals, different problems in which some local, time varying degree of freedom interacts with a fermionic environment, can be treated with the same methodology and hence some incarnation of the fidelity amplitude and Loschmidt echo appears naturally. Examples include the absorption spectrum of Luttinger liquids~\cite{Schotte:1969,Nagaosa:1992,
Glazman:1994} and beyond~\cite{Imambekov:2012}, the single channel Kondo problem~\cite{Yuval:1970} or time-dependent impurites in cold atom systems~\cite{Knap:2012}.

In order to make the connection with the Langevin equation discussed above, in this section we obtain an expression for the Loschmidt echo {\it via} a perturbative expansion for small changes in the potential $\mathcal{V}_{{\bf X}}$, by considering $\mathcal{H}_\mathrm{i}=\mathcal{H}_0  + \mathcal{V}_{{\bf X}}$ and $\mathcal{H}_\mathrm{f}=\mathcal{H}_0  + \mathcal{V}_{{\bf X}+ \delta{\bf X}}$, cf. Fig.~\ref{fig:mov_scatt}, with $\delta \mathcal{H}_{\mathbf{X}} =\mathcal{V}_{\mathbf{X} + \delta \mathbf{X}} - \mathcal{V}_{\mathbf{X}}$ small with respect to  $\mathcal{H}_\mathrm{i}$. An important factor to determine the time dependence of the Loschmidt echo, is {\it how rapidly} the change in the coupling potential occurs. This rapidity is determined by what is called the ``quench protocol''. Here we consider this is given externally by an arbitrary function $g(t)$ such that
\begin{align}\label{eq:TDH}
\mathcal{H}(t) = \mathcal{H}_0  + \mathcal{V}_{{\bf X}}+ g(t) \, \delta \mathcal{H}_{\mathbf{X}} 
\end{align}
 Initially, $g(0)=0$ so that we obtain $\mathcal{H}_\mathrm{i}$. We impose the quench is completed at some time $\tau$ by setting $g(\tau) = 1$. We consider an open quantum system in which the electrons spend on average some finite time $\tau_D$ in the scattering region. $\tau_D$ is refered to as the {\it dwell time} and we consider it to be the smallest time scale in the system, in the spirit of the Born-Oppenheimer approximation. This defines the time scale for the quench. In particular in this work we will study two complementary protocols: sudden and adiabatic quenches. In the {\it sudden quench}, the scattering potential is changed suddenly, which is realized by a step-function shape such that $g(t) = 1$ for $t > 0$. For the {\it adiabatic quench} instead, the potential is ramped up slowly, where slow refers to the dwell time. For the adiabatic quench, we choose a linear ramping protocol $g(t) = t/\tau$. We show in Appendix~\ref{app:AdiabaticProtocols} that, in the limit $\tau\gg\tau_D$, our results in equilibrium 
are 
independent of this choice, while for an imposed bias our results are characterized by coefficients that depend on the specifics of the adiabatic protocol. In what follows we will treat in parallel both abovementioned protocols, we list in Table~\ref{tab:steuerzeichen} the protocol-dependent parameters as used in the text.
\begin{table}[t]
\centering  \label{tab:steuerzeichen}
\begin{tabular}{c|c|c}
symbol & Sudden ($P=S$) & Adiabatic ($P=A$)\\\hline
$\lambda_P$ & $1$ & $2$ \\ 
 $\alpha_P$ & $2$ & $1$ \\
 $\beta_P$ & $1$ & $1/3$ \\
 $\delta_P$ & $2$ & $1/2$ \\
\end{tabular}
\caption{Protocol-dependent constants used throughout the text. Note that while $\alpha_A$ is universal and valid for any adiabatic quench, the rest of the ``adiabatic'' constants are valid exclusively for the linear quench --- they are actually dependent on the adiabatic quench protocol.}
\end{table} 
The Hamiltonian defined in Eq.~\eqref{eq:TDH} is time dependent through $g(t)$. To treat this time dependence we write the fidelity amplitude in terms of evolution operators
\begin{align}\label{eq:FA_U}
\mathcal{A}(\tau) = \langle  \, U^\dagger_0(\tau,0) \, U(\tau,0)   \, \rangle \,.
\end{align}
The operator $U_0$ is the time-evolution operator of the (constant) initial Hamiltonian $\mathcal{H}_0 + \mathcal{V}_{\mathbf{X}}$ and $U$ that of the (time-dependent) Hamiltonian $\mathcal{H}(t)$. In the case of a sudden quench, we recover the usual expression Eq.~\eqref{eq:FA_def}. We introduce now the interaction picture with respect to the initial Hamiltonian. This allows us to write
\begin{align}\label{eq:FA_IP}
 \mathcal{A}(\tau)  = \langle \hat{T} \, \exp \Big( - i \int\limits_0^{\tau} \textrm{d}t\, g(t) \, \delta \mathcal{H}_{\mathbf{X}}(t)    \Big)  \, \rangle 
\end{align}
with time-ordering operator $\hat{T}$ and $\delta\mathcal{H}_{\mathbf{X}}(t) = e^{i \mathcal{H}_{\rm{i}} t} \delta\mathcal{H}_{\mathbf{X}}  e^{-i \mathcal{H}_{\rm{i}} t} $. The expression given in Eq.~\eqref{eq:FA_IP} is the starting point for the perturbative expansion given below.

We now perform a pertubative expansion of the fidelity amplitude in the displacement $\delta \mathbf{X}$ up to the first non-zero terms both in imaginary and real parts. We assume the scattering potential to be well behaved, such that small changes in ${\bf X}$ correspond to small changes in  $\mathcal{V}$.~\footnote{In a scattering approach, ``small changes in the potential'' can be understood as $\tau_D\delta\mathcal{H}_{\bf X}\ll 1$, since the electronic Hamiltonian is of the order of $1/\tau_D$. For the adiabatic quench the change in the potential is time dependent such that most of the scattering electrons see a change smaller than $\delta\mathcal{H}_{\bf X}$, hence the condition is also satisfied.} The corresponding change in the Hamiltonian is
\begin{align}
\delta \mathcal{H}_{\mathbf{X}} = \sum_{\alpha} \partial_\alpha \mathcal{V}_{\mathbf{X}} \delta X_\alpha = \sum_{\alpha} \partial_\alpha \mathcal{H}_{\mathbf{X}} \delta X_\alpha\,.
\end{align}
We therefore obtain
\begin{align}\label{eq:FA_PE}
\begin{split}
  \ln \mathcal{A}(\tau) &= - i \int\limits_0^\tau \textrm{d}t \, g(t) \, \langle  \delta\mathcal{H}_{\mathbf{X}}(t)  \rangle  \\
  & -   \frac{1}{2} \int\limits_0^\tau \textrm{d}t \int\limits_0^\tau \textrm{d}t'  \sum_{\alpha \beta}  g(t) \, g(t') \,  D_{\alpha\beta}(t,t')   \delta X_\alpha \, \delta X_\beta \,,
\end{split}
\end{align}
where $D_{\alpha\beta}(t,t')$ is the noise correlator given in Eq.~(\ref{eq:ColorNoise}) and we have used Eqs.~\eqref{eq:ForceOperator} and~\eqref{eq:LangevinForce}. As anticipated in the previous section, due to the sum over the indices $\alpha$ and $\beta$, only the symmetric component of the noise correlator is relevant. The second term in Eq.~(\ref{eq:FA_PE}) is a real quantity, while the first order term is purely imaginary and hence contributes as an overall phase. This phase is directly related to the infinitesimal work made by the Born-Oppenheimer force, which is consistent with the shift in the dynamical phase of the system's eigenstates which is acquired due to the change in the potential $\delta \mathcal{H}_{\mathbf{X}}$. This can be seen by expressing the quantum statistical average  $\langle\delta \mathcal{H}_{\mathbf{X}}\rangle$ in terms of scattering states.~\cite{Thomas:2012} For the adiabatic ($P=A$) and sudden ($P=S$) quenches we obtain 
\begin{align}\label{eq:FA_phase_AQ}
 \mathcal{A}_{P}(\tau) = e^{i \,\frac{\tau}{\lambda_P}  \, \mathbf{F}(\mathbf{X}) \cdot  \delta\mathbf{X}}   |\mathcal{A}_{P}(\tau)| \,,
\end{align}
with $\lambda_A=2$ and $\lambda_S=1$, where the reduction by a factor of $2$ of the adiabatic fidelity phase with respect to the sudden quench results from integrating over the linear ramp-up of the potential.

The Loschmidt echo, in turn, is given solely in terms of the integrated two-time noise correlation function
\begin{align}\label{eq:LEvariance}
 \ln  \mathcal{L}(\tau)   = - \int\limits_0^\tau \textrm{d}t \int\limits_0^\tau \textrm{d}t' g(t) \, g(t') \,  \mathbf{\delta X}^\dagger\cdot \mathbf{D}(t,t')  \cdot \mathbf{\delta X}\,,
\end{align}
where the function $g(t)$ enters as a weight factor. We note that assuming Gaussian white noise, where $\mathbf{D}(t,t')$ is delta-correlated in time, it immediately follows from Eq.~(\ref{eq:LEvariance}) that the Loschmidt echo decays exponentially with a strength proportional to $\mathbf{\delta X}^\dagger \cdot\mathbf{D}(\mathbf{X})  \cdot \mathbf{\delta X}$ in the large time limit. For now we keep the results general and discuss the regime of applicability for the white noise limit later. Expressing the fluctuating force correlator by its Fourier transform we readily observe that the decay of the Loschmidt echo is determined by the symmetric noise correlator $\mathbf{D}(\omega)$,~\footnote{Note that $\mathbf{D}(\omega)$ is not the Fourier transform of $\mathbf{D}(t-t')$. It is rather the symmetrized Fourier transform (in frequency~$\omega$).}
\begin{align}\label{eq:LE_and_D_omega}
   \ln  \mathcal{L}(\tau)   = -  \int\limits_0^{\infty} \frac{\textrm{d}\omega}{\pi} \left| \, \int\limits_0^\tau \textrm{d}t    \,g(t) \, e^{i \omega t} \, \right|^2  \mathbf{\delta X}^\dagger \cdot  \mathbf{D}(\omega) \cdot \mathbf{\delta X}\,.
\end{align}
The time integral above can be performed once the quench dynamics $g(t)$ is specified 
\begin{align}\label{eq:LE_PQ}
  \ln  \mathcal{L}_{P}(\tau) = -\int\limits_0^{\infty} \frac{\textrm{d}\omega}{\pi}\,B_P(\omega,\tau)\,\mathbf{\delta X}^\dagger \cdot  \mathbf{D}(\omega) \cdot \mathbf{\delta X}\,,
\end{align}
where the function $B_P(\omega,\tau)$ is protocol-dependent 
\begin{align}
 \begin{split}
 B_S(\omega,\tau)= &2\,\frac{1 - \cos (\omega \, \tau )}{ \omega^2 }\\
 B_A(\omega,\tau)= & \frac{2 \, \left[1-\cos (\omega\,\tau)-\omega\tau\sin (\omega\tau)\right] +\omega^2\tau^2 } {\tau^2\omega^4}  \,.
 \end{split}
\end{align}
The dwell time $\tau_D$ is a characteristic time scale for the scattering of the fast (electronic) degrees of freedom and provides a high energy cutoff $1/\tau_D$ for the energy integrals in Eq.~(\ref{eq:LE_PQ}). At the same time, the function $B_P(\omega,\tau)$ selects frequencies $\omega\lesssim 1/\tau$. This allows us, in the limit of large $\tau\gg\tau_D$, to neglect the dynamics of the fast degrees of freedom and evaluate the fluctuating force correlator $\mathbf{D}(\omega)$ in Eq.~\eqref{eq:LE_PQ} in the limit of small frequencies, $\omega \sim 1/\tau \sim 0$ --- we will use this fact in the next sections.

The expressions obtained in this section are, within the limit of validity of the perturbative approach, quite general. In particular, they hold for out-of-equilibrium situations. To investigate these results we start in the next section with the equilibrium case, for which we can straightforwardly apply the fluctuation-dissipation theorem as given in Eq.~\eqref{eq:fluct-diss}. The out-of-equilibrium regime is treated later in Sec.~\ref{sec:nonequilibrium}. 

\section{Equilibrium}\label{sec:Equilibrium}

We consider here the equilibrium case for which all leads are kept at a same chemical potential denoted by $\mu$. In equilibrium the Anderson orthogonality exponent has been shown in Ref.~\onlinecite{Schoenhammer:1991} to be proportional to the friction coefficient of the noninteracting fermionic environment for finite systems. This corresponds to the $\tau \rightarrow \infty$ limit of the Loschmidt echo.~\cite{Muender:2012} We generalize this result here to the case of an open system with a continuous energy spectrum by calculating the decay of the Loschmidt echo for finite times $\tau$. 

We continue with evaluating Eq.~(\ref{eq:LE_PQ}) at $T=0$ and we discuss the effect of finite temperature later. We can therefore use the fluctuation-dissipation theorem as given in Eq.~\eqref{eq:FD_Q}. We conclude that, to leading order in $\tau/\tau_D$,
\begin{align}\label{eq:LE_fluct-diss}
   \ln  \mathcal{L}_{P}(\tau)= - \frac{1}{\pi}\! \int\limits_0^{1/\tau_D}\!\! \omega\,\textrm{d}\omega\,B_{P}(\omega,\tau)\,  \mathbf{\delta X}^\dagger \cdot\boldsymbol{\gamma}^{\mathrm{\rm{eq}}} \cdot \mathbf{\delta X}\,,
\end{align}
where $\boldsymbol{\gamma}^{\mathrm{\rm{eq}}}$ is the equilibrium friction cofficient evaluated at zero frequency. We see therefore that, as expected, in equilibrium the Loschmidt echo is closely related to dissipation. We obtain
\begin{align}\label{eq:LE_SQ_equilibrium}
 \ln  \mathcal{L}_{P}(\tau) = - \, \frac{\alpha_P}{\pi} \left[\ln \left( \frac{\tau}{\tau_D} \right) + \gamma_e \right]   \mathbf{\delta X}^\dagger \cdot\boldsymbol{\gamma}^{\mathrm{\rm{eq}}}(\mathbf{X}) \cdot \mathbf{\delta X}
\end{align}
where $\alpha_P$ is protocol dependent, with $\alpha_S=2$, $\alpha_A=1$ for the sudden and adiabatic quenches respectively, and $\gamma_e = 0.5772$ is the Euler-Mascheroni constant. We show in App.~\ref{app:AdiabaticProtocols}, that the value $\alpha_A=1$ is independent of the assumption of linearity for the adiabatic quench protocol. Therefore,
\begin{align}\label{eq:SQ_AQ_equilibrium}
 \mathcal{L}_{P}(\tau)  \propto \left( \frac{\tau}{\tau_D}   \right)^{- \, \frac{\alpha_P}{\pi} \mathbf{\delta X}^\dagger \cdot\boldsymbol{\gamma}^{\mathrm{\rm{eq}}}(\mathbf{X}) \cdot \mathbf{\delta X}} \,,
\end{align}
so that the decay of the Loschmidt echo in equilibrium, both in the sudden and adiabatic quench scenarios, is a powerlaw controlled by the friction coefficient of the fermionic system. The powerlaw decay of the Loschmidt echo is consistent with known literature results\cite{Dominicis:1969,Schoenhammer:1991} and reflects the Anderson orthogonality catastrophe.~\cite{Anderson:1967} For a finite system the $\tau\rightarrow \infty$ limit of the Loschmidt echo can be obtained, up to prefactors, by 
replacing 
$\tau/\tau_D$ by the number of particles of the system and the powerlaw takes the usual Anderson's form. This is justified since the ratio $\tau/\tau_D$ can be taken as an estimate of how many particles have been scattered up to time $\tau$ --- for $\tau\rightarrow \infty$, all particles in the system have participated in the scattering.

Note that the powerlaw decay of the Loschmidt echo is inconsistent with a delta-correlated noise --- recall that white noise implies an {\it exponential} decay of the Loschmidt echo. The powerlaw decay signals the breakdown of the Markovian, semi-classical Langevin equation~\eqref{eq:Langevin} in equilibrium and at zero temperature, for which case the classical noise correlator is zero. In other words, the system loses its memory as a powerlaw in time instead of exponentially, which renders the Markovian approximation inapplicable. An exponential decay of the Loschmidt echo is recovered either by imposing finite temperature or a finite bias voltage. In the following we comment on the finite temperature case, and we reserve the next section for out-of equilibrium effects.   

For temperatures such that $T\gg1/\tau$, we can use the classical version of the fluctuation-dissipation theorem Eq.~\eqref{eq:FD_CL} in Eq.~\eqref{eq:LEvariance} to obtain
\begin{align}
 \ln \mathcal{L}_P(\tau) = - 2\,\beta_P \,\tau\,T \delta\mathbf{X}^\dagger
\cdot \boldsymbol{\gamma}^{\mathrm{\rm{eq}}} \cdot  \delta \mathbf{X}\,,
\end{align}
and therefore we recover, for $\tau \gg \tau_D$, an exponential decay governed by the thermal noise 
\begin{align}\label{eq:LexpT}
 \mathcal{L}_P(\tau) =e^{ -  \tau \,  \beta_P \, \delta\mathbf{X}^\dagger
\cdot \mathbf{D}_{[T,0]} \cdot  \delta \mathbf{X}}
\end{align}
where $\beta_P$ is protocol dependent, with $\beta_S=1$, $\beta_A=1/3$ for the sudden and adiabatic (linear) quench respectively. Note that in this case the coefficient $\beta_A$ depends on the nature of the adiabatic protocol.

As a last remark of this section, we observe that in equilibrium and zero temperature the adiabatic and sudden Loschmidt echo are related by a simple exponent.~\footnote{Note that these relations are valid for $\tau/\tau_D \gg 1$.} For zero temperature, from Eq.~\eqref{eq:LE_SQ_equilibrium} we obtain
\begin{align}\label{eq:LA_vs_LS}
 \mathcal{L}_{S}(\tau)  = \mathcal{L}_{A}(\tau)^2\,.
\end{align}
As pointed out before, $\alpha_A=1$ is independent of the adiabatic protocol, and therefore Eq.~\eqref{eq:LA_vs_LS} holds generally. The relation given by Eq.~\eqref{eq:LA_vs_LS} has been recently pointed out for particular examples in Refs.~~\onlinecite{Zarand:2013, Sachdeva:2013} for infinite~$\tau$ in finite systems, and argued to be valid in more general situations.~\cite{Zarand:2013}  For finite temperatures we obtain instead from Eq.~\eqref{eq:LexpT}
\begin{align}\label{eq:LA_vs_LS_T}
 \mathcal{L}_{S}(\tau)  = \mathcal{L}_{A}(\tau)^{1/\beta_A}\,,
\end{align}
which is valid to leading order in $\tau_D/\tau$.

\section{Out-of-equilibrium}\label{sec:nonequilibrium}

In this section we take a step further and allow for the presence of an applied bias voltage, represented by different chemical potentials in the leads. For clarity we consider only two leads which are kept at a chemical potential difference $\Delta \mu>0$. We obtain the decay of the Loschmidt echo in the limit of linear response, for which the applied bias is small as defined by the condition $\Delta\mu \,\tau_D\ll1$. By evaluating Eq.~(\ref{eq:LE_PQ}) with the out-of-equilibrium noise correlator given in Eq.~\eqref{eq:D_omega_sym_tot}, we can obtain closed expressions for the time dependence of the Loschmidt echo in terms of the macroscopic coefficients appearing in the Langevin Eq.~\eqref{eq:Langevin}. These expressions are valid for all times longer than the dwell time, as detailed below. The derivation is lenghty and therefore we summarize here the main results and give a sketch of the calculation in the next 
subsections, while the details can be found in the corresponding appendices as listed.

It is instructive to consider the long- and short-time dynamics of the Loschmidt echo as compared with the timescale determined by the inverse of the imposed bias, since in these limits the expressions simplify considerably. We state here the results for zero temperature. The case of {\it short-time dynamics} (but still large times compared with the dwell time) is given by the condition $\Delta\mu\,\tau\ll1$. For these short times the system is being probed at high energies and it is not sensitive to the applied bias. We therefore recover the equilibrium result
\begin{align}\label{eq:LE_EQ_intro}
 \mathcal{L}_{P}(\tau) \propto  \left( \frac{\tau}{\tau_D}  \right)^{- \frac{\alpha_P}{\pi}\mathbf{\delta X}^\dagger\cdot\boldsymbol{\gamma}\cdot\mathbf{\delta X}}\,,
\end{align} 
Here $\boldsymbol{\gamma}$ is the full dissipation matrix evaluated to first order in the bias. Ignoring this first order correction due to the bias we recover exactly the equilibrium result obtained previously in Eq.~\eqref{eq:LE_SQ_equilibrium}. 

In the opposite limit of {\it long-time dynamics}, $\Delta\mu\,\tau \gg 1$, the system is more sensitive to the non-equilibrium imposed by the bias which results in a different qualitative behavior. The major effect due to bringing the system out of equilibrium is an exponential suppression of the Loschmidt echo in the long-time dynamics, compared with the equilibrium power-law decay in Eq.~\eqref{eq:LE_SQ_equilibrium}. We obtain 
\begin{align}\label{eq:LE_NE_intro}
\begin{split}
\mathcal{L}_{P}(\tau) \,\propto\, & e^{-\,\beta_P \, \tau \mathbf{\delta X}^\dagger\cdot\mathbf{D}_{[0,\Delta\mu]}\cdot\mathbf{\delta X} }\left(\Delta\mu\tau_D\right)^{\frac{\alpha_P}{\pi} \mathbf{\delta X}^\dagger\cdot\frac{\mathbf{D}_{[0,\Delta\mu]}}{\Delta\mu}\cdot\mathbf{\delta X}}\\
\times&   \left(\frac{\tau}{\tau_D}\right)  ^{- \frac{\alpha_P}{\pi} \mathbf{\delta X}^\dagger\cdot\left[\boldsymbol{\gamma}^{\mathrm{\rm{eq}}}-\frac{\mathbf{D}_{[0,\Delta\mu]}}{\Delta\mu}\right]\cdot\mathbf{\delta X}} \,,
\end{split}
\end{align}
where the dissipation matrix and noise correlators are evaluated to first order in the bias. The exponential suppression of the Loschmidt echo is dictated by the shot-noise fluctuations in the system given by the noise correlator $\mathbf{D}_{[0,\Delta\mu]}$ to first order in the bias.~\cite{Aleiner:1997} Note that this exponential decay is completely analogous to that for equilibrium and finite temperatures in the long-time limit ($T\tau\gg1$) given in Eq.~\eqref{eq:LexpT}, which is dominated by the termal noise.

The exponential decay in Eq.~\eqref{eq:LE_NE_intro} comes on top of a powerlaw behavior, with an exponent that is also modified from equilibrium showing a competition between fluctuations and dissipation, and with the possibility of a change of sign in the exponent. This powerlaw correction is absent in the complementary case of equilibrium and finite temperatures given in Eq.~\eqref{eq:LexpT}, due to the vanishing of the linear order in frequency correction for the thermal noise correlator in Eq.~\eqref{eq:FD_CL}. Note that the out-of-equilibrium powerlaw crosses over to the equilibrium one at a time $\tau\approx 1/\Delta\mu$ as expected.
 
We proceed now with the derivation of these results.

\subsection{Out-of-equilibrium noise correlator}
We see from Eq.~\eqref{eq:LEvariance} that the force-force noise correlator is crucial to determine the behavior of the Loschmidt echo. In this section we derive the expression given in Eq.~\eqref{eq:D_omega_sym_tot} for the noise correlator within a scattering approach which highlights the connection to the current-induced forces in the Langevin Eq.~\eqref{eq:Langevin}. An equivalent derivation in terms of Keldysh Green's functions is given, for completeness, in App.~\ref{app:Keldysh}. Alternatively, the correlator can be calculated within a Feynman Vernon influence functional approach.~\cite{Lue:2012} 

We now proceed with evaluating $\mathbf{D}(t,t')$ as given in Eq.~\eqref{eq:ColorNoise} in terms of single-particle scattering states. For this we introduce the notation
\begin{align}\label{eq:Ham_scat_states}
\begin{split}
\partial_\alpha V^{kn}_{\mathbf{X}}(\varepsilon,\varepsilon') = \langle \psi^{\mathbf{X}+}_{k}(\varepsilon) | \partial_\alpha V_{\mathbf{X}} | \psi^{\mathbf{X} +}_{n}(\varepsilon') \rangle 
\end{split}
\end{align}
for the matrix elements of the representation of $\nabla\mathcal{V}_{\mathbf{X}}$ in the scattering basis --- cf. App.~\ref{app:scattering1}. Here $ | \psi^{\mathbf{X} +}_{n}(\varepsilon) \rangle$ is the single-particle retarded scattering state with combined channel-lead index $n$ and energy $\varepsilon$. For notational convenience in what follows, we further define the function
\begin{align}\label{eq:def_F}
K^{\alpha\beta}_{kn}(\varepsilon,\varepsilon') =   \left\{\partial_\alpha V^{kn}_{\mathbf{X}}(\varepsilon,\varepsilon') \, \partial_\beta V^{nk}_{\mathbf{X}}(\varepsilon',\varepsilon)\right\}_\mathrm{\rm{s}} \,.
\end{align}
Using Eq.~\eqref{eq:many_body} to evaluate the quantum statistical expectation values~\cite{Buttiker:1992} appearing in the noise correlator~(\ref{eq:ColorNoise}) we obtain
\begin{align}\label{eq:noise_D}
 \begin{split}
 \mathbf{ D}(t,t') &= \int\frac{\textrm{d}\varepsilon}{2\pi}\int\frac{\textrm{d}\varepsilon'}{2\pi} \sum_{kn}  f_k(\varepsilon) \, [1 - f_n(\varepsilon')] \\
  & \times \,  e^{i(\varepsilon - \varepsilon' ) \,(t - t ') } \, \mathbf{ K}_{kn}(\varepsilon,\varepsilon')  \vphantom{\int} \,.
 \end{split}
\end{align}
After Fourier transforming we obtain a general expression for the force-force noise correlator as a function of frequencies~\cite{Levinson:2000}
\begin{align}\label{eq:D_omega_gen}
\begin{split}
 \mathbf{ D}(\omega) &= \int\frac{\textrm{d}\varepsilon}{2\pi} \sum_{kn} f_k\left(\varepsilon - \frac{\omega}{2}\right) \, \left[1-f_n\left(\varepsilon + \frac{\omega}{2}\right)\right] \\
&\times\mathbf{ K}_{kn}\left(\varepsilon - \frac{\omega}{2} ,\varepsilon + \frac{\omega}{2}\right)\,.
\end{split}
\end{align}
We observe that the function $\mathbf{K}_{kn}(\varepsilon,\varepsilon')$ contains overlaps of scattering states which are are associated with scattering events including an energy transfer $\omega = \varepsilon -\varepsilon' $. We expect these overlaps to vary within energies up to the inverse dwell time $1/\tau_D$. We can expect our description of the Loschmidt echo in terms of scattering states to be valid in the limit $\tau \gg \tau_D$ --- a description on microscopic time scales smaller than $\tau_D$ is beyond an adiabatic scattering formulation. Hence in the following we restrict our calculations to this limit, and evaluate $\mathbf{K}_{kn}(\varepsilon,\varepsilon')$ to first order in $\omega\ll1/\tau_D$. We will see later in the text that this is enough to capture the leading behavior of the Loschmidt echo as a function of time. It is also convenient to define the function 
\begin{align}
\mathbf{K}_{kn}(\varepsilon) = \mathbf{K}_{kn}(\varepsilon,\varepsilon)\,.
\end{align}
We note that the function $\mathbf{K}_{kn}(\varepsilon)$ is closely related to the dissipation matrix in equilibrium. At zero temperature and taking all leads to be at an equal chemical potential $\mu$, as we show in Appendix~\ref{app:scattering2}, it takes the simple form
\begin{align}\label{eq:gamma0_relation}
  \frac{1}{4\pi} \sum_{kn}\mathbf{K}_{kn}(\mu)=\boldsymbol{\gamma}^{\mathrm{\rm{eq}}}\,,
\end{align}
in agreement with the result found previously by use of the fluctuation-dissipation theorem.

The second contribution to Eq.~\eqref{eq:D_omega_gen} is given by the product of Fermi functions $f_k(\varepsilon) [1 - f_n(\varepsilon')]$. Due to this product, the average energy $\overline{\varepsilon} = (\varepsilon + \varepsilon')/2$ is limited to a region of size $\Delta\mu_{kn} = \mu_k - \mu_n$ around the respective average chemical potential $\overline{\mu}_{kn}=1/2(\mu_k+\mu_n)$. To make analytical progress, we limit our results to the linear response regime $\Delta\mu_{kn} \, \tau_D \ll 1\,$, which allows a perturbative treatment of the function $ \mathbf{K}_{kn}(\varepsilon,\varepsilon')$ for small deviations of $\overline{\varepsilon}$ around $\overline{\mu}_{kn}$. 

Given these considerations, we calculate Eq.~\eqref{eq:D_omega_gen} to leading order in $\tau_D/\tau$, in the linear response regime (linear order in $\Delta\mu\,\tau_D$). Expanding  $\mathbf{K}_{kn}(\varepsilon,\varepsilon')$ to first order in $\omega$ and $\overline{\varepsilon}-\overline{\mu}_{kn}$ we obtain
\begin{align}\label{eq:K_Exp} 
\begin{split}
\mathbf{K}_{kn}(\varepsilon,\varepsilon')&=  \mathbf{K}_{kn}(\overline{\mu}_{kn}) \\
&+ 2\,( \overline{\varepsilon} -  \overline{\mu}_{kn})  \, \partial^{\rm{s}}_{\varepsilon} \mathbf{K}_{kn}(\overline{\mu}_{kn})  +   \omega \,  \partial^{\rm{a}}_{\varepsilon} \mathbf{K}_{kn}(\overline{\mu}_{kn}),
\end{split}
\end{align}
where we have introduced 
\begin{align}
 \partial^{\rm{s}/{a}}_\varepsilon \mathbf{K}_{kn}(x) =  \frac{1}{2} \,\left( \partial_{\varepsilon} \pm \partial_{\varepsilon'} \right) \,\mathbf{K}_{kn}(\varepsilon,\varepsilon') \Big|_{\varepsilon'=\varepsilon=x} \,,
\end{align}
which describes the symmetric and antisymmetric energy derivatives of $\mathbf{K}_{kn}$. We note in passing the following useful properties: $\mathbf{K}_{kn}(\varepsilon) = \mathbf{K}_{nk}(\varepsilon)$, $\partial^{\rm{s}}_{\varepsilon} \mathbf{K}_{kn}(\varepsilon) = \partial^{\rm{s}}_{\varepsilon}\mathbf{K}_{nk}(\varepsilon)$ and $\partial^{\rm{a}}_{\varepsilon} \mathbf{K}_{kn}(\varepsilon) = - \partial^{\rm{a}}_{\varepsilon} \mathbf{K}_{nk}(\varepsilon)$. Substituting Eq.~\eqref{eq:K_Exp} into Eq.~\eqref{eq:D_omega_gen}, the energy integral can be performed to obtain
\begin{align}
 \begin{split}
 \mathbf{D}(\omega) &= \frac{1}{2\pi} \sum_{kn} \,\frac{\omega + \Delta \mu_{kn}}{e^{(\omega + \Delta \mu_{kn})/T} - 1} \\
&\times e^{(\omega + \Delta \mu_{kn})/T}  \,\left[  \mathbf{K}_{kn}(\overline{\mu}_{kn})  - \omega \, \partial^{\rm{a}}_{\varepsilon} \mathbf{K}_{kn}(\overline{\mu}_{kn}) \right] \,.
 \end{split}
\end{align}
To be consistent with the linear response approximation, the functions $\mathbf{K}_{kn}$ and $\partial^{\rm{s}/{a}}_\varepsilon \mathbf{K}^{\alpha\beta}_{kn}$ have to be evaluated to first order in $\Delta\mu \tau_D$. We proceed with this expansion below for the case of two leads, for which the expressions are more transparent. The indices $k,n$ describe hereafter (the two) lead indices only, where we implicitly assume a summation over the channel index. A generalization to an arbitrary number of leads is straightforward. Without loss of generality we write $\mu_R = \mu - \Delta\mu/2$, $\mu_L = \mu + \Delta\mu/2$ with $\Delta\mu > 0$. Explicitly, $\mathbf{K}_{LL}(\mu_L) = \mathbf{K}_{LL}(\mu) + \Delta\mu \partial^{\rm{s}}_{\varepsilon} \mathbf{K}_{LL}(\mu)$ and $\mathbf{K}_{RR}(\mu_R) = \mathbf{K}_{RR}(\mu) - \Delta\mu \partial^{\rm{s}}_{\varepsilon} \mathbf{K}_{RR}(\mu)$. Hence, within linear response we obtain (with $\omega_\pm=\omega\pm\Delta\mu$)
\begin{widetext}
\begin{align}\label{eq:Noise_Tot_Exp}
  \mathbf{D}(\omega)&= \frac{1}{4\pi} \,\omega \, \coth \left(\frac{\omega}{2\,T} \right) \, \left[  \mathbf{K}_{LL}(\mu) + \mathbf{K}_{RR}(\mu) + \Delta\mu \left( \partial^{\rm{s}}_{\varepsilon} \mathbf{K}_{LL}(\mu) - \partial^{\rm{s}}_{\varepsilon} \mathbf{K}_{RR}(\mu) \right)   \right] \\
  &+ \frac{1}{4\pi} \left[ \omega_+  \, \coth \left( \frac{\omega_+}{2\,T} \right) + \omega_-  \, \coth \left( \frac{\omega_-}{2\,T} \right) \right] \, \mathbf{K}_{LR}(\mu) \\
  &+ \frac{\omega}{4\pi} \left[ - \omega_+  \, \coth \left( \frac{\omega_+}{2\,T} \right) + \omega_-  \, \coth \left( \frac{\omega_-}{2\,T} \right) \right] \, \partial^{\rm{a}}_{\varepsilon} \mathbf{K}_{LR}(\mu) \,.
\end{align}
\end{widetext}
The coefficients appearing in the expansion can be interpreted in terms of the different dissipative contributions with the help of the relations found in App.~\ref{app:scattering1}. The connection to the friction tensor is found by expanding the dissipation tensor $\boldsymbol{\gamma}$ to first order in $\Delta \mu \tau_D$ as~\footnote{Note that the linear response expansion is indeed an expansion in $\Delta\mu\,\tau_D$ since the energy derivatives appearing in Eqs.~\eqref{eq:gamma_1eq} and~\eqref{eq:gamma_1neq} are of order the Wigner time delay, which 
can be estimated to be of the order of $\tau_D$.} $\boldsymbol{\gamma}= \boldsymbol{\gamma}^{\mathrm{\rm{eq}}}_0 + \boldsymbol{\gamma}^{\mathrm{\rm{eq}}}_1+\boldsymbol{\gamma}^{\mathrm{\rm{neq}}}_1+ \ldots$, 
where the subscript $0$ ($1$) denotes the zeroth (first) order in the expansion respectively of the equilibrium ($\mathrm{eq}$) and non-equilibrium ($\mathrm{neq}$) contributions to the friction tensor. In App.~\ref{app:scattering2} we show the following identities
\begin{align}\label{eq:gamma_1eq}
\boldsymbol{\gamma}_1^{\rm{eq}}  &= \frac{\Delta\mu}{4 \pi} \left[ \partial^{\rm{s}}_{\varepsilon} \mathbf{K}_{LL}(\mu) -      \partial^{\rm{s}}_{\varepsilon} \mathbf{K}_{RR}(\mu)\right] \,,\\\label{eq:gamma_1neq}
\boldsymbol{\gamma}_1^{\rm{neq}} &=  \frac{\Delta\mu}{4 \pi} \left[ \partial^{\rm{a}}_{\varepsilon} \mathbf{K}_{RL}(\mu) -      \partial^{\rm{a}}_{\varepsilon} \mathbf{K}_{LR}(\mu)\right]\,\\\label{eq:D_neq}
\mathbf{D}_{[0,\Delta\mu]} &=\frac{\Delta\mu}{4 \pi} \, \left[ \mathbf{K}_{LR}(\mu) + \mathbf{K}_{RL}(\mu)\right]\,,
\end{align}
which together with Eq.~\eqref{eq:gamma0_relation} imply
\begin{align}\label{eq:aux_gamma_D}
 \mathbf{K}_{LL}(\mu) + \mathbf{K}_{RR}(\mu)= 4\pi \left(\boldsymbol{\gamma}_0^{\rm{eq}} - \frac{\mathbf{D}_{[0,\Delta\mu]}}{\Delta\mu} \right)\,.
\end{align}
Plugging in these identities in Eq.~\eqref{eq:Noise_Tot_Exp} we obtain the anticipated result stated in Eq.~\eqref{eq:D_omega_sym_tot}.

In order to be sensitive to non-equilibrium effects, we need to impose temperatures smaller than the bias. We therefore take the zero temperature limit of the noise correlator given in Eq.~\eqref{eq:D_omega_sym_tot}. For $|\omega| < \Delta\mu$ we then obtain
\begin{align}\label{eq:D_omega_sym1} 
\mathbf{ D}(\omega) = \mathbf{D}_{[0,\Delta\mu]} + |\omega| \, \left(  \boldsymbol{\gamma}^{\mathrm{\rm{eq}}}   - \frac{\mathbf{D}_{[0,\Delta\mu]}}{\Delta \mu} \right) + \frac{\omega^2}{2} \frac{\boldsymbol{\gamma}^{\mathrm{\rm{neq}}}}{\Delta\mu}\,,
\end{align}
while for $|\omega|>\Delta\mu$ we have
 \begin{align}\label{eq:D_omega_sym2} 
\mathbf{ D}(\omega) = |\omega| \boldsymbol{\gamma}\,,
\end{align}
where we have used that $\lim_{x\to \pm\infty} \, \coth x = \pm 1$.

\subsection{Short- and long-time dynamics}\label{sec:LE_3}
Inserting Eqs.~\eqref{eq:D_omega_sym1} - \eqref{eq:D_omega_sym2} into Eq.~\eqref{eq:LE_PQ} we  can express the decay of the Loschmidt echo in terms of the mesoscopic coefficients that control the Langevin dynamics of a heavy particle embedded in the fermionic environment, cf. Eq.~\eqref{eq:Langevin}. We obtain
\begin{align}\label{eq:LE_long_time_ext_intermediate}
\begin{split}
 \ln &\mathcal{L}_{P}(\tau) =- \, \frac{\alpha_P}{\pi} \, \left[ \gamma_e + \ln\left(\frac{\tau}{\tau_D} \right) \right]\delta\mathbf{X}^\dagger\cdot\boldsymbol{\gamma} \cdot \delta \mathbf{X} \\
 &+   \frac{\alpha_P}{\pi} \, \Big[ \gamma_e + \ln\left(\Delta\mu \tau \right) \Big] \delta\mathbf{X}^\dagger\cdot \left( \frac{\mathbf{D}_{[0,\Delta\mu]}}{\Delta\mu} + \boldsymbol{\gamma}_1^{\rm{neq}} \right)\cdot \delta \mathbf{X}\\
&- \beta_P \, \tau  \, \delta \mathbf{X}^\dagger\cdot\mathbf{D}_{[0,\Delta\mu]}\cdot \delta \mathbf{X} -  \frac{\delta_P}{\pi} \, \delta\mathbf{X}^\dagger\cdot  \boldsymbol{\gamma}_1^{\rm{neq}} \cdot \delta \mathbf{X} \\
& + \, \frac{1}{\pi}  \Big(  \alpha_P - 2 \,\beta_P \, \cos(\Delta\mu\,\tau) \Big)\, \delta \mathbf{X}^\dagger\cdot \frac{\mathbf{D}_{[0,\Delta\mu]}}{\Delta\mu}\, \cdot \delta \mathbf{X}\,.
\end{split}
\end{align}
with $\delta_S = 2$, $\delta_A = 1/2$. Eq.~\eqref{eq:LE_long_time_ext_intermediate} gives the behavior of the Loschmidt echo at arbitrary times larger than the dwell time, up to quadratic order in $\delta \mathbf{X}$. In the following we further investigate different timescale regimes.

The short-time dynamics is given by the limit $\Delta\mu \, \tau \ll 1$ --- ``short times'' here should be considered as short with respect to the inverse bias time scale but long compared to the dwell time $\tau_D$. In this regime we conclude 
\begin{align}\label{eq:LE_short_time_ext}
 \ln \mathcal{L}_{P}(\tau) =  - \, \frac{\alpha_P}{\pi} \left[ \gamma_e + \ln\left(\frac{\tau}{\tau_D} \right) \right]\delta\mathbf{X}^\dagger\cdot  \boldsymbol{\gamma} \cdot \delta \mathbf{X}\,,
\end{align}
which yields Eq.~\eqref{eq:LE_EQ_intro}. Note that the equilibrium friction term given by $\boldsymbol{\gamma}_0^{\rm{eq}}$ constitutes the dominant contribution for the decay, and the full friction matrix $\boldsymbol{\gamma}$ is restricted to positive values for small $\Delta\mu\,\tau_D \ll 1$, which ensures that $ \mathcal{L}_{PQ}(\tau)\leq 1$ for all times $\tau$. 

We turn now to evaluating Eq.~\eqref{eq:LE_long_time_ext_intermediate} in the long-time limit $\Delta\mu \, \tau \gg 1$. Writing $\ln(\tau/\tau_D)=\ln(\Delta\mu\tau/(\Delta\mu\tau_D))$ and observing that $\Delta\mu\tau_D \, \ln(\Delta\mu \tau_D)$ goes to zero for $\Delta\mu\tau_D \ll 1$ we obtain
\begin{align}\label{eq:LE_long_time_ext}
\begin{split}
 \ln &\mathcal{L}_{P}(\tau)=- \beta_P \, \tau  \, \delta \mathbf{X}^\dagger\cdot\mathbf{D}_{[0,\Delta\mu]}\cdot \delta \mathbf{X} \\
&- \frac{\alpha_P}{\pi} \,\Big[ \gamma_e + \ln\left(\Delta\mu \tau \right) \Big] \delta\mathbf{X}^\dagger\cdot\left(  \boldsymbol{\gamma}^{\mathrm{\rm{eq}}}  - \frac{\mathbf{D}_{[0,\Delta\mu]}}{\Delta\mu} \right)\cdot\delta\mathbf{X}   \\
 &+ \frac{\alpha_P}{\pi} \, \ln\left(\Delta\mu \tau_D \right)   \delta \mathbf{X}^\dagger\cdot \boldsymbol{\gamma}_0^{\rm{eq}}\cdot \delta \mathbf{X}\\
&-  \frac{\delta_P}{\pi} \, \delta\mathbf{X}^\dagger\cdot  \boldsymbol{\gamma}_1^{\rm{neq}} \cdot \delta \mathbf{X} \\
& + \, \frac{1}{\pi}  \Big(  \alpha_P - 2 \,\beta_P\, \cos(\Delta\mu\,\tau) \Big)\, \delta \mathbf{X}^\dagger\cdot \frac{\mathbf{D}_{[0,\Delta\mu]}}{\Delta\mu}\, \cdot \delta \mathbf{X}\,,
\end{split}
\end{align}
from which we obtain Eq.~\eqref{eq:LE_NE_intro} by keeping the dominant terms in large~$\tau$.~\cite{Muzykantskii:2003}

The exponential suppression of the Loschmidt echo is dictated by the shot-noise fluctuations in the system (note that $\mathbf{D}_{[0,\Delta\mu]}$ is positive definite) and it is consistent with a leading behavior of Gaussian white noise for the fluctuating force correlator $\mathbf{D}(t,t')$ in Eq.~(\ref{eq:ColorNoise}). Furthermore, the exponent of the powerlaw  shows a competition between fluctuations and dissipation, which is a clear signature of the departure from equilibrium --- as remarked before, the corresponding powerlaw is absent in the equilibrium, finite temperature case. Since $\boldsymbol{\gamma}^{\mathrm{\rm{eq}}}=\mathbf{D}_{[T,0]}/(2T)$ for zero bias, the sign of the exponent $-\delta\mathbf{X}^\dagger\cdot [\boldsymbol{\gamma}^{\mathrm{\rm{eq}}}-\mathbf{D}_{[0,\Delta\mu]}/(\Delta \mu)]\cdot\delta\mathbf{X}$ depends on the assymetry between shot and Nyquist noise in the linear response regime. In fact, this exponent can be positive for finite bias, which leads to an enhancement of the powerlaw instead of the usual decay, which has been dubbed as ``anti-orthogonality''~\cite{Segal:2007}. We note however that $- 
\delta\mathbf{X}^\dagger\cdot [\boldsymbol{\gamma}_0^{\rm{eq}}-\mathbf{D}_{[0,\Delta\mu]}/(\Delta \mu)]\cdot\delta\mathbf{X}$ is always negative (see Eqs.~\eqref{eq:aux_gamma_D} and \eqref{eq:K_def_app}). The sign change of the exponent happens when the leading order $\mathbf{K}_{RR}(\mu) \approx - \mathbf{K}_{LL}(\mu)$ cancels out, such that the linear order correction $\boldsymbol{\gamma}_1^{\rm{eq}}$ in the bias becomes dominant, which can then lead to a change of the sign of the exponent. %We show this behavior in the next section for a specific model.

Since the out-of-equilibrium short-time dynamics is essentially the equilibrium one, the identity $\mathcal{L}_{S}(\tau)  = \mathcal{L}_{A}(\tau)^2$ is still fullfilled in this limit, as can be seen directly from Eq.~\eqref{eq:LE_short_time_ext}. On the other hand, in the large-time regime, we conclude from Eq.~\eqref{eq:LE_long_time_ext} that this identity is violated due to the factor $\beta_P$ in the exponential. This difference, attributed to the structure of $g(t)$, can already be obtained by looking at Eq.~(\ref{eq:LEvariance}). Assuming white noise, we immediately deduce from Eq.~(\ref{eq:LEvariance}) an exponential decay of the Loschmidt echo with an exponent $- \beta_P \delta\mathbf{X}^\dagger\cdot \mathbf{D}_{[0,\Delta\mu]}\cdot \delta\mathbf{X} \, \tau $. The powerlaw decay in Eq.~\eqref{eq:LE_long_time_ext} constitutes minor correction terms to the white noise assumption, and therefore to leading order in $\tau_D/\tau$ we obtain $\mathcal{L}_{S}(\tau)  = \mathcal{L}_{A}(\tau)^{1/\beta_A}$ as in the equilibrium, finite temperature case given in Eq.~\eqref{eq:LA_vs_LS_T}.

In Sec.~\ref{sec:2-level_model} we study these results for a specific example of a two-level model coupled to one vibrational mode. 

\section{Example: two-level model with one vibrational mode}\label{sec:2-level_model}
In this section we analyze the sudden quench Loschmidt echo for the example of a system with one classical degree of freedom connected to two leads. This serves as an toy model to illustrate the above results. The ``heavy'' classical degree of freedom $X= X(t)$ corresponds to a mechanical vibrational mode of the system. Accordingly we consider the Hamiltonian
\begin{align}\label{eq:Niels_Hamiltonian}
\mathcal{H}= \mathcal{H}_X + \mathcal{H}_L + \mathcal{H}_D + \mathcal{H}_T 
\end{align}
where the different terms are specified as
\begin{align}
\mathcal{H}_X &= \frac{P^2}{2M} + U(X)   \\
\mathcal{H}_L&= \int \frac{\textrm{d}\varepsilon}{2\pi}\sum_{\eta}\left(\varepsilon-\mu_{\eta}\right)c_{\eta}^{\dagger}(\varepsilon)c_{\eta}(\varepsilon) \\
\mathcal{H}_D &= \sum_{mm'}d_{m}^{\dagger}\left[h_0(X)\right]_{mm'}d_{m'} \\\label{eq:Niels_Hamiltonian_4}
\mathcal{H}_T &= \int \frac{\textrm{d}\varepsilon}{\sqrt{2\pi}}\sum_{\eta m}\left(c_{\eta}^{\dagger}(\varepsilon)W_{\eta m}(\varepsilon)d_{m}+h.c.\right) \,.
\end{align}
Here, the operator $c_{\eta}^{\dagger}(\varepsilon)$ [$c_{\eta}(\varepsilon)$] creates [annihilates] electronic states $|\phi_\eta(\varepsilon)\rangle$, which are approaching the scattering region from lead $\eta = L, R$ with chemical potential~$\mu_L \geq \mu_R$. $\mathcal{H}_X$ describes the evolution of the parameter $X$ with potential $U(X)$, mass $M$ and frequency $\omega_0$. $H_D$ models the two-level system (quantum dot) with states $|m\rangle$, created (annihilated) by the operators $d_{m}^{\dagger}$ ($d_{m}$). $H_T$ represents tunneling between the leads and the system with tunneling amplitudes $W_{\eta m}(\varepsilon)=\langle \phi_\eta(\varepsilon)|W|m\rangle/\sqrt{2\pi}$. The coupling of the mechanical degree of freedom and the electrons in the dot is described by the matrix~$h_0(X)$.

\begin{figure}
		\includegraphics[width=0.35\textwidth]{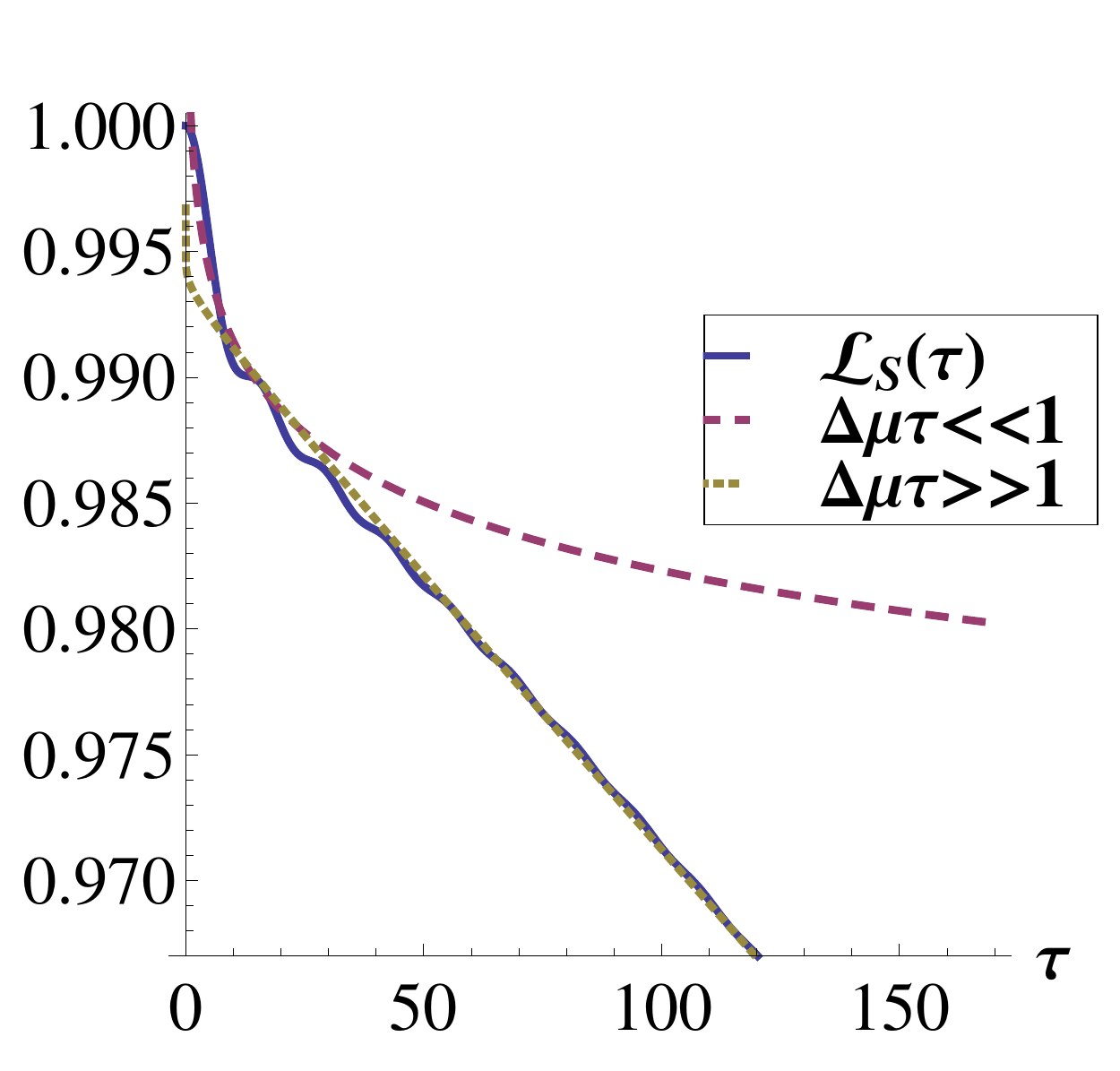}
		\caption{Sudden quench Loschmidt echo as given in Eq.~\eqref{eq:LE_long_time_ext_intermediate} and comparison to the short- and long-time behavior Eqs.~\eqref{eq:LE_EQ_intro},\eqref{eq:LE_NE_intro} (with the corresponding proportionality coefficients), for the example of a two-level system coupled to one vibrational mode with $\Gamma = t = 0.48$, $\tau_D = 1/\Gamma = 2.08$, $\varepsilon_0=0$, $\Delta \mu = 0.04$, $\delta X = 0.15$, $X = 0.4$, $T=0$. All energies (and inverse times) are in units of $\lambda^2/(M \omega_0^2)$ and distances in units of $\lambda/(M \omega_0^2)$.}\label{fig:1-level_LE_plot}
\end{figure}

We consider a two-level system with degenerate energy levels $\varepsilon_0$. The single oscillator mode $X$ is assumed to couple to the difference in the energy level occupation with a strength given by $\lambda$. Hence we write 
\begin{align}
 h_0(X) = \left( \begin{array}{cc}
                    \varepsilon_+ & t \\
                    t & \varepsilon_-
                   \end{array} \right).
\end{align}
with interdot tunneling amplitude $t$ and $\varepsilon_{\pm} = \varepsilon_0  \pm \lambda \, X$. Tunneling from the left (right) lead to the two-level system and back is described by the amplitudes $\Gamma_L$ ($\Gamma_R$) which for simplicity we take as $\Gamma_L = \Gamma_R = \Gamma/2$. In the wide-band approximation these amplitudes are assumed to be energy independent. With these definitions, the coupling matrix $W_{\eta m}$ reads
\begin{align}
 W =      \left( \begin{array}{cc}
                    \sqrt{\Gamma/(2\pi)}  & 0 \\
                    0 & \sqrt{\Gamma/(2\pi)}
                   \end{array} \right)  \,,
\end{align}
and the frozen retarded Green's function takes the form
\begin{align}
 G^R_{X}(\varepsilon) = \frac{1}{\Delta_X(\varepsilon)} \,
		    \left( \begin{array}{cc}
                    \varepsilon - \varepsilon_-  + i\,\Gamma/2 &  t \\
                    t & \varepsilon - \varepsilon_+  + i\,\Gamma/2
		    \end{array} \right) 
\end{align}
with $\Delta_X(\varepsilon)=  (\varepsilon - \varepsilon_-  + i\,\Gamma/2) \, (\varepsilon - \varepsilon_+  + i\,\Gamma/2) - t^2$. This model was studied in Ref.~\onlinecite{Bode:2012} in the context of current-induced forces. The frozen scattering matrix $S$ and its first order non-adiabatic correction $A$ are given by
\begin{align}\label{eq:example_S-Matrix}
 S_{X}(\varepsilon) = \mathbbm{1} - \frac{i\,\Gamma}{L_X(\varepsilon)} 
 \left( \begin{array}{cc}
   1 & 1 \\
   1 & 1
   \end{array}
 \right) \,,
\end{align}
\begin{align}
A_{X}(\varepsilon) = \frac{\lambda \, \Gamma \, t}{\Delta_X(\varepsilon)^2} \left( \begin{array}{cc}
   0 & 1 \\
   -1 & 0
   \end{array}\right) \,,
\end{align}
where $L_X(\varepsilon) = \varepsilon - \varepsilon_+ + i \, \Gamma$. With the expression of the S-matrix and the A-matrix, we can determine all the mesoscopic coefficients appearing in the Langevin Eq.~(\ref{eq:Langevin}), which determine the behavior of the Loschmidt echo. This is depicted for arbitrary times $\tau>\tau_D$ in Fig.~\ref{fig:1-level_LE_plot} according to Eq.~\eqref{eq:LE_long_time_ext_intermediate}, and compared to the small and large time regimes expressions in Eqs.~\eqref{eq:LE_EQ_intro},\eqref{eq:LE_NE_intro}. The dwell time enters in this example as a timescale which is of the order of the inverse tunneling amplitudes.

\begin{figure}
		\subfigure[]{\includegraphics[width=0.23\textwidth]{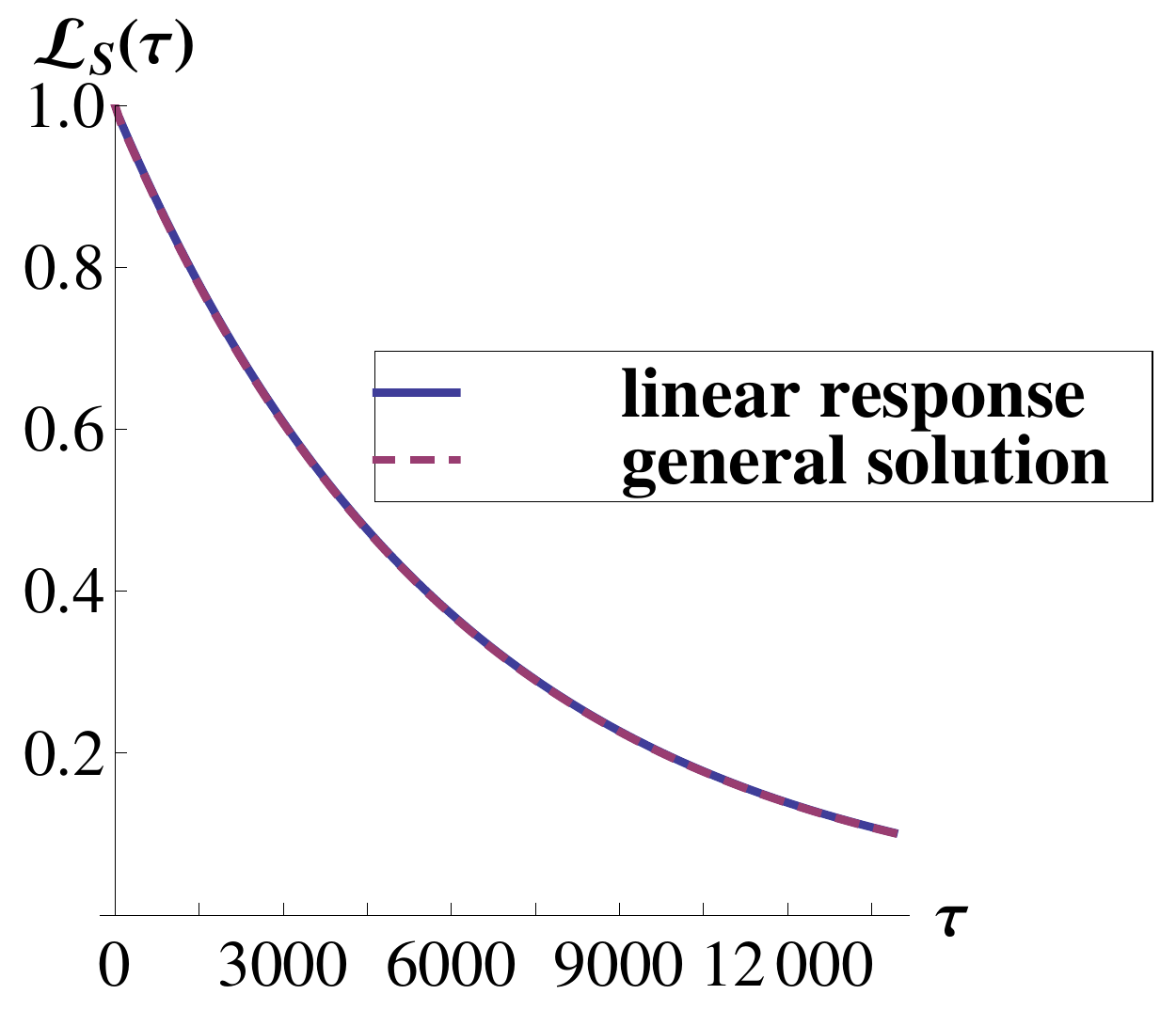}}
		\subfigure[]{\includegraphics[width=0.23\textwidth]{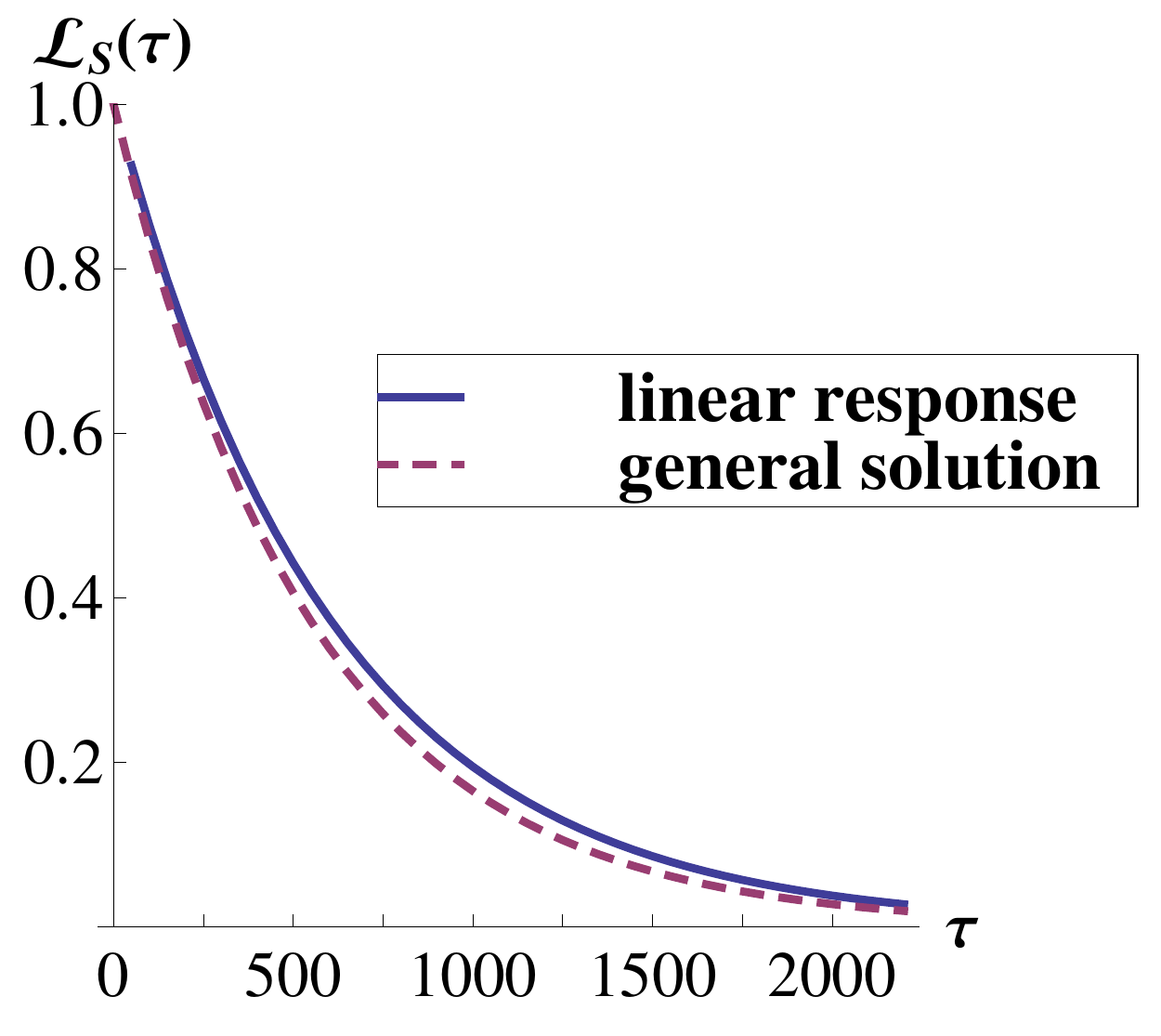}}
		\caption{Linear response solution of the sudden Loschmidt echo --- Eq.~\eqref{eq:LE_long_time_ext_intermediate}, vs. general solution [i.e. Eq.~(\ref{eq:D_omega_gen}) inserted into Eq.~(\ref{eq:LE_PQ})] at different bias voltages, (a) $\Delta \mu = 0.04$, (b) $\Delta \mu = 0.4$; others parameters as in Fig.~\ref{fig:1-level_LE_plot}.}\label{fig:1-level_lin_resp_vs_gen}
\end{figure}
The considered model allows us to analyze the Loschmidt echo also outside of the linear response regime, by directly evaluating the colored noise-noise correlator as given in Eq.~(\ref{eq:D_omega_gen}). The matrix $\partial_\alpha V_{\mathbf{X}}(\varepsilon,\varepsilon')$ [see Eq.~(\ref{eq:Ham_scat_states})] is given by
\begin{align}
  \partial_\alpha V^{kn}_{\mathbf{X}}(\varepsilon,\varepsilon') = 2 \pi\left[ W \cdot G^R_{X}(\varepsilon)^\dagger   \partial_X h_0(X)  G^R_{X}(\varepsilon') \cdot W^{\dagger} \right]_{kn}.
\end{align}
Eq.~(\ref{eq:D_omega_gen}) is neither restricted to $\Delta\mu \, \tau_D \ll 1$ nor to the regime $\tau_D/\tau \ll 1$ and hence is valid for arbitrary $\Delta \mu$ and all $\tau$. Thus we can study the Loschmidt echo for increasing bias voltages $\Delta \mu$. A comparison of the general solution, that is substituting Eq.~(\ref{eq:D_omega_gen}) into Eq.~(\ref{eq:LE_PQ}), and the linear response solution Eq.~\eqref{eq:LE_long_time_ext_intermediate}, is depicted in Fig.~\ref{fig:1-level_lin_resp_vs_gen}. The figure shows that the linear response solution agrees very well with the general solution for small~$\Delta\mu \, \tau_D$.

\begin{figure}
		\subfigure[]{\includegraphics[width=0.23\textwidth]{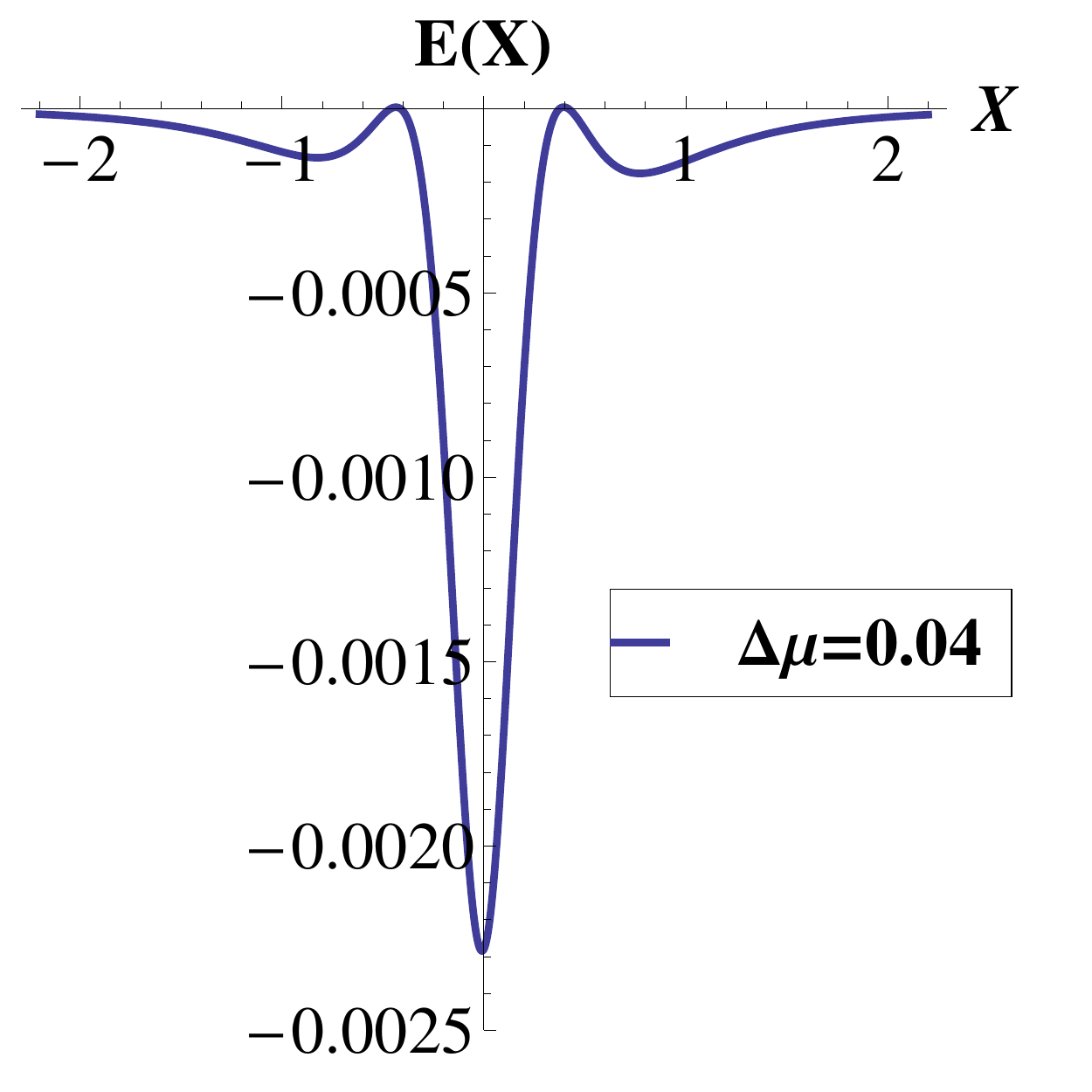}}
		\subfigure[]{\includegraphics[width=0.23\textwidth]{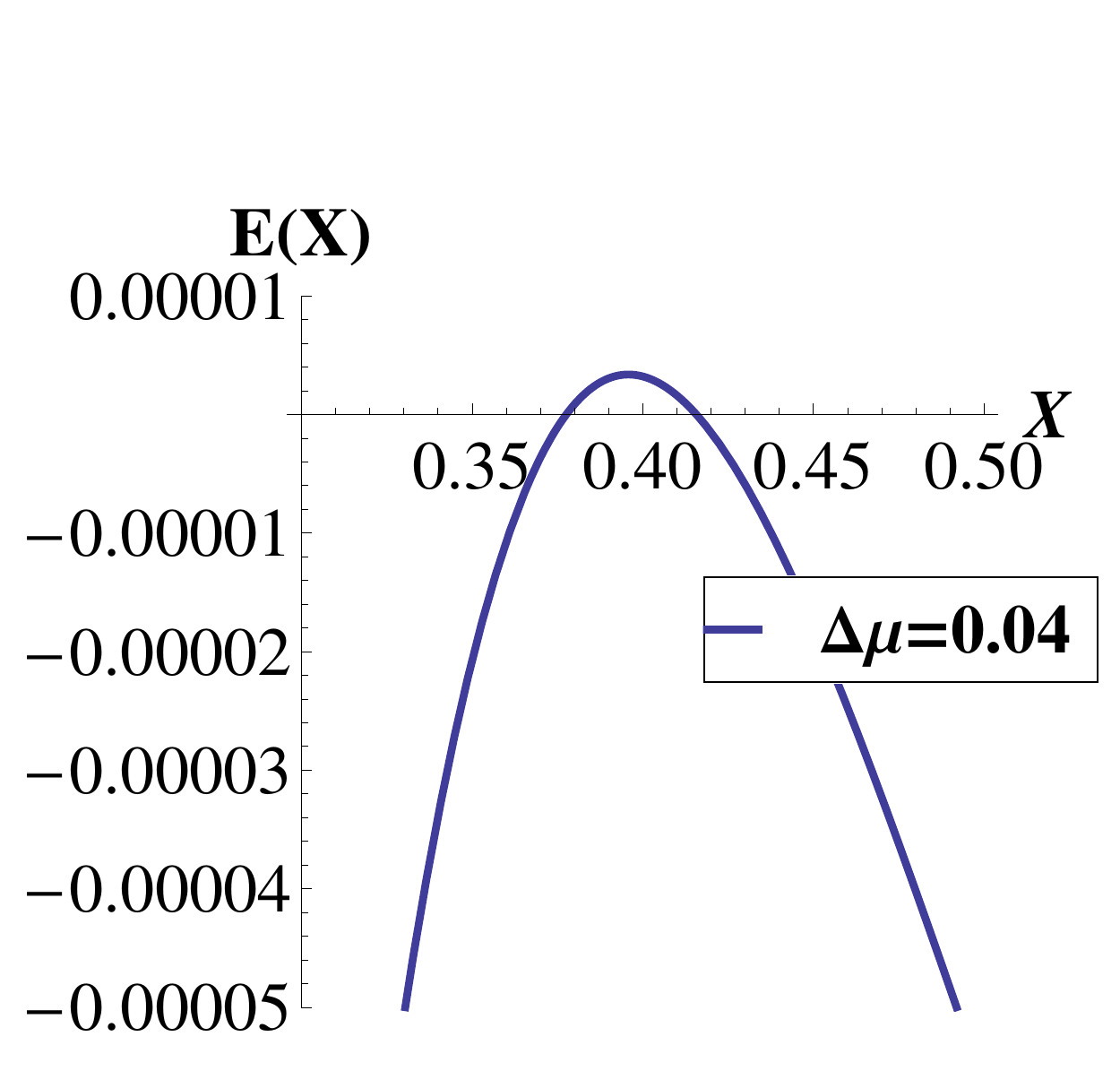}}
		\caption{(a) The out-of-equilibrium powerlaw coefficient ${\rm{E}}(X)=- \frac{2}{\pi} (\gamma_0^{\rm{eq}} + \gamma_1^{\rm{eq}} - \frac{D}{\Delta\mu})$  for $\Delta\mu = 0.04$ as a function of $X$; other parameters as in Fig.~\ref{fig:1-level_LE_plot}. (b) Zoom-in of (a) which shows the change of sign in the exponent.}\label{fig:anti-orthogonality}
\end{figure}

To close this section, we show that the powerlaw exponent $- \frac{2}{\pi} (\boldsymbol{\gamma}^{\rm{eq}} - \mathbf{D}_{[0,\Delta\mu]}/\Delta \mu)$ can present changes in sign. This is shown in Fig.~\ref{fig:anti-orthogonality} for a specific finite bias voltage, where the exponent becomes positive for a small range of displacements. 

\section{Summary}\label{sec:conclusion}

In this paper we calculated perturbatively the decay of the Loschmidt echo for an open non-interacting fermionic system in the presence of an external bias, which is subject to a scattering potential quench. We expressed our results in terms of the mesoscopic quantities describing the complementary problem of a heavy particle moving in a quantum environment, and showed that the Loschmidt echo decay is controlled by the noise correlator of the heavy particle. 

This result allowed us to study Anderson orthogonality effects for the open quantum system. In the particular case of equilibrium and zero temperature, the decay of the Loschmidt echo is a power law controlled by the dissipation coefficient of the heavy particle, in agreement with the results found in Ref.~\onlinecite{Schoenhammer:1991} in which the exponent of the Anderson
orthogonality was related to the dissipation of a heavy particle
moving in a finite-sized quantum environment. For finite temperatures, in the limit of long times we recovered an exponential decay which reflects the classical version of the
fluctuation-dissipation theorem.

When a small bias is imposed, we showed that in the long-time dynamics (as compared to the energy scale given by the bias), the Loschmidt echo is dictated by an
exponential decay with a strength given by the shot-noise fluctuations. The exponential decay is consistent with a white-noise spectrum to leading order. As a correction term to white noise, the Loschmidt echo shows an algebraic behavior with a powerlaw
exponent given by a competition between fluctuations and dissipation. This competition can give rise to changes in the sign of the powerlaw exponent. The powerlaw correction to the exponential decay is characteristic of non-equilibrium effects and is absent in equilibrium at finite temperatures. In the case of short-time dynamics, the system is mostly insensitive to the applied bias and the Loschmidt echo still presents a powerlaw decay, controlled by the full non-equilibrium dissipation coefficient. 

The results summarized above are independent of the quench scenario. In particular we studied the complementary cases of sudden and slow quenches, and showed that in equilibrium the sudden quench Loschmidt echo is the square of the Loschmidt echo for the slow quench, independent of the functional form of the slow quench. This
relation generalizes the relation found for finite quantum systems at infinite times in Ref.~~\onlinecite{Zarand:2013, Sachdeva:2013}, where a Luttinger liquid model which is subject to a linear slow quench was studied. We find that this correspondence breaks down out of equilibrium or for finite temperatures. To leading order however we can still establish a simple relation involving a non-universal exponent.

%%%%%%%%%%%%%%%%%%%%%%%%%%%%%%%%%%%%%%%%%%%%%%%%%%%%%%
\section*{Acknowledgements}
We thank Felix von Oppen for several illuminating discussions. We acknowledge support by the Deutsche Forschungsgesellschaft through Sfb 658 (S.V.K., M.T.), the Helmholtz Virtual
Institute ``New States of Matter and Their Excitations''
(Berlin) and the A. v H. Foundation (S.V.K), and the Institute for Quantum Information and Matter, a NSF Physics Frontiers Center with support of the Gordon and Betty Moore Foundation through Grant GBMF1250 (T.K.).

%%%%%%%%%%%%%%%%%%%%%%%%%%%%%%%%%%%%%%%%%%%%%%%%%%%%%%

\begin{appendices}
 \numberwithin{equation}{section} %for equation numbering in the appendix as A.1, A.2 etc.
\begin{widetext}
\section{Adiabatic Protocols}\label{app:AdiabaticProtocols}

In this appendix we show that Eq.~(\ref{eq:SQ_AQ_equilibrium}) is independent of the adiabatic quenching protocol. In particular we show that for any function $g(t)$ which grows as a powerlaw to its maximal value on a timescale much larger than~$\tau_D$, we obtain a powerlaw decay of the Loschmidt echo as in Eq.~(\ref{eq:LE_SQ_equilibrium}) for $\alpha_P = 1$. We are interested in the equilibrium behaviour at zero temperature. Equation~\eqref{eq:LE_fluct-diss} is valid for any quenching protocol with $g(0)=0$ and $g(\tau) = 1$. Specifying the quench dynamics to $g(t) = \left( \frac{t}{\tau} \right)^n $ with $n \geq 1$ integer we read off
\begin{align}\label{eq:powerlaw_protocol}
  \ln \mathcal{L}_{AQ}(\tau) = - \frac{1}{\pi} \int\limits_0^{1/\tau_D} \textrm{d}\omega \, \omega \, \left| \int_0^\tau\textrm{d}t \,\left( \frac{t}{\tau} \right)^n e^{i\omega t} \right|^2\mathbf{\delta X}^\dagger \cdot\boldsymbol{\gamma}^{\mathrm{\rm{eq}}}(\mathbf{X}) \cdot \mathbf{\delta X}\,.
\end{align}
We split the integral with respect to $\omega$ into a part $\omega \leq 1/\tau$ and $\omega \geq 1/\tau$. Since the integral where $\omega \leq 1/\tau$ gives only a constant (which depends on the value of $n$), it is irrelevant at large $\tau/\tau_D \gg 1$. We are thus left with
\begin{align}
  \ln \mathcal{L}_{AQ}(\tau) = - \frac{1}{\pi}\int\limits_{1/\tau}^{1/\tau_D} \textrm{d}\omega \, \omega \, \left| \int_0^\tau\textrm{d}t \,\left( \frac{t}{\tau} \right)^n e^{i\omega t} \right|^2 \mathbf{\delta X}^\dagger \cdot\boldsymbol{\gamma}^{\mathrm{\rm{eq}}}(\mathbf{X}) \cdot \mathbf{\delta X}\,.
\end{align}
We perform an integration by parts to evaluate the integral in the absolute square values. For large $\tau/\tau_D$ we approximate to leading order
\begin{align}
  \int_0^\tau\textrm{d}t \,\left( \frac{t}{\tau} \right)^n e^{i\omega t} = \frac{e^{i \omega \tau}}{i\omega} \,.
\end{align}
Inserting this into Eq.~(\ref{eq:powerlaw_protocol}) we conclude that
\begin{align}
\begin{split}
  \ln \mathcal{L}_{AQ}(\tau) = - \frac{1}{\pi} \sum_{\alpha \beta} \, \ln\left(\frac{\tau}{\tau_D} \right)  \mathbf{\delta X}^\dagger \cdot\boldsymbol{\gamma}^{\mathrm{\rm{eq}}}(\mathbf{X}) \cdot \mathbf{\delta X}
\end{split}
\end{align}
up to an irrelevant constant for all $g(t) = \left( \frac{t}{\tau} \right)^n $ with $n \geq 1$ integer. Comparing to the large time behaviour of the Loschmidt echo for the sudden quench scenario  we readily deduce the validity of Eq.~(\ref{eq:SQ_AQ_equilibrium}) independent of the adiabatic protocol.

\section{Adiabatic scattering theory}\label{app:scattering1}

%%%%%%%%%%%%%%%%%%%%%%%%%%%%%%%%%%%%%%5

In this appendix we outline the elements of scattering theory used in the main text. For $\mathcal{H}_{\mathbf{X}} = \mathcal{H}_0 + \mathcal{V}_\mathbf{X}$ assuming non-interacting particles, we can write
\begin{align}\label{eq:many_body}
 \mathcal{H}_{\mathbf{X}} = \int\frac{\textrm{d}\varepsilon}{2\pi}\int\frac{\textrm{d}\varepsilon'}{2\pi} \sum_{kn} [H_{\mathbf{X}}]_{kn}(\varepsilon,\varepsilon')  a^{\mathbf{X} \dagger}_k(\varepsilon)  a^{\mathbf{X}}_n(\varepsilon')  \,,
\end{align}
with the \emph{single-particle} Hamiltonian $ H_{\mathbf{X}} = H_0 + V_{\mathbf{X}}$. The operators $a^{\mathbf{X}}_m(\varepsilon)$ and $a^{\mathbf{X}\dagger}_m(\varepsilon)$ annihilate and create, respectively, the retarded single-particle scattering states $|\psi^{\mathbf{X} +}_m(\varepsilon)\rangle$ of the Hamiltonian $H_{\mathbf{X}}$ with energy $\varepsilon$ and combined channel and lead index $m$, hence it follows
\begin{align}\label{eq:H_mn}
[H_{\mathbf{X}}]_{kn}(\varepsilon,\varepsilon') = \langle \psi^{\mathbf{X} +}_{k}(\varepsilon) | H_{\mathbf{X}}| \psi^{\mathbf{X} +}_{n}(\varepsilon') \rangle\,. 
\end{align}
The corresponding advanced scattering states are indicated with the superscript~($-$), that is $ | \psi_m^{\mathbf{X} -} (\varepsilon) \rangle $. These scattering states are solutions of the time-independent Schr\"odinger equation at every time $t$ [note that we consider $\mathbf{X}=\mathbf{X}(t)$], and obey the Lippmann-Schwinger equation. The retarded and advanced scattering states are defined by their boundary conditions: $ | \psi_m^{\mathbf{X} +(-)} (\varepsilon) \rangle $ has incoming (outgoing) waves only in channel/lead $m$, and evolved from the free states $ | \phi_m(\varepsilon)\rangle  $ at $t \rightarrow \mp \infty$, which fulfill $H_0 \,| \phi_m(\varepsilon)\rangle = \varepsilon \, | \phi_m(\varepsilon)\rangle$ with energy $\varepsilon$. We normalize the scattering states to unit flux. The overlap between retarded and advanced scattering states then defines the S-matrix as
\begin{align}\label{eq:S-matrix}
S_{\mathbf{X}}^{lm}(\varepsilon) \, 2\pi \, \delta(\varepsilon-\varepsilon')  = \langle \psi_l^{\mathbf{X} -} (\varepsilon)  | \psi_m^{\mathbf{X} +} (\varepsilon') \rangle \,. 
\end{align}
From this definition it follows that $S_{\mathbf{X}}^\dagger(\varepsilon) S_{\mathbf{X}}(\varepsilon) = 1$, because of the scattering states' normalization to unit flux.

For a Hamiltonian which parametrically depends on time via the classical parameter~$\mathbf{X}$, Eq.~\eqref{eq:S-matrix} gives the frozen S-matrix of the system: the solution of the time-independent scattering problem at time~$t$. The {\it exact} S-matrix~$\mathcal{S}(\varepsilon,\varepsilon')$ is defined through the overlap of retarded and advanced scattering states solution of the full time-dependent problem. Assuming a slowly varying parameter~$\mathbf{X}$, the exact S-matrix can be expressed as an adiabatic expansion in the velocity $\dot{\mathbf{X}}_t$~\cite{Moskalets:2004,Arrachea:2006,Bode:2011,Bode:2012,Thomas:2012}. In this context, ``slowly varying'' means that the dynamics of $\mathbf{X}$ is much slower than the electronic time scales. To first order in the adiabatic expansion we can write, in the Wigner representation, $\mathcal{S}(\varepsilon,t) = S_{\mathbf{X}}(\varepsilon) + \sum_\alpha A_{\mathbf{X}}^\alpha(\varepsilon)  \dot{X}_\alpha$. The zeroth order is given 
by the frozen S-matrix. The first non-
adiabatic correction to the S-
matrix is given by~\cite{Thomas:2012}
\begin{align}\label{eq:A-matrix}
A^{lm, \alpha}_{\mathbf{X}}(\varepsilon)  = \frac{1}{2} \, \langle \partial_{\varepsilon} \psi_l^{\mathbf{X} -} (\varepsilon)  | \,\partial_{\alpha}V_{\mathbf{X}} \, |\psi_m^{\mathbf{X} +} (\varepsilon) \rangle    - \frac{1}{2} \, \langle  \psi_l^{\mathbf{X} -} (\varepsilon)  | \,\partial_{\alpha}V_{\mathbf{X}} \, |  \partial_{\varepsilon}\psi_m^{\mathbf{X} +} (\varepsilon) \rangle    \,. 
\end{align}
with $\partial_{\alpha} = \partial/\partial X_\alpha $ and $|  \partial_{\varepsilon}\psi_m^{\mathbf{X} \pm} (\varepsilon) \rangle \equiv   \partial_{\varepsilon} | \psi_m^{\mathbf{X} \pm} (\varepsilon) \rangle$. Note that throughout the manuscript we work in the Heisenberg picture, so that there is no explicit time dependence on the states. due to unitarity, the S and A matrix fulfill~\cite{Thomas:2012} 
\begin{align}
  \langle  \psi_n^{\mathbf{X}-}(\varepsilon) | \partial_\alpha V |  \psi_k^{\mathbf{X}+}(\varepsilon) \rangle &=  i\, \partial_\alpha S_{nk}(\varepsilon)  \,,\\
 \langle  \psi_n^{\mathbf{X}-}(\varepsilon) | \partial_\alpha V | \partial_{\varepsilon} \psi_k^{\mathbf{X}+}(\varepsilon) \rangle &= -A^{\alpha}_{nk}(\varepsilon)  + \frac{i}{2} \, \partial_\varepsilon \partial_\alpha S_{nk}(\varepsilon)\,.
 \end{align}
These relations will be used in the next appendix.

We can express the reaction forces in the Langevin Eq.~\eqref{eq:Langevin} in terms of Eqs. \eqref{eq:S-matrix} and \eqref{eq:A-matrix}, which we list here for completeness.~\cite{Bode:2011,Bode:2012,Thomas:2012} The Born-Oppenheimer force is given by 
\begin{align}\label{eq:Born-Oppenheimer_force}
 F_{\alpha}(\mathbf{X}) = \int \frac{\textrm{d}\varepsilon}{2\pi i} \sum_n f_n(\varepsilon)\textrm{tr}\left\{\Pi_n S_{\mathbf{X}}^{\dagger}(\varepsilon)\partial_\alpha S_{\mathbf{X}}(\varepsilon) \right\}\,,
\end{align}
where $\textrm{tr}\{\ldots\} $ denotes a trace over scattering channels, $f_n(\varepsilon) = [\exp[(\varepsilon-\mu_n)/T + 1]]^{-1}$ is the fermionic distribution function in lead $n$ with chemical potential $\mu_n$, and $\partial_\alpha = \partial/\partial X_\alpha$, $\Pi_n$ is a projector onto channel~$n$. The two contributions to the dissipative force as discussed in the main text are in turn given by
 \begin{align}\label{eq:EQ_friction}
  \gamma_{\alpha \beta}^{\rm{eq}}(\mathbf{X}) &= - \sum_{n}\int\frac{\textrm{d}\varepsilon}{4\pi}  \partial_\varepsilon[ f_n(\varepsilon)]\textrm{tr} \left\{ \Pi_n \partial_\alpha S_{\mathbf{X}}^{\dagger}(\varepsilon)  \partial_\beta S_{\mathbf{X}}(\varepsilon)  \right\}_\mathrm{\rm{s}} \\
  \gamma_{\alpha \beta}^{\rm{neq}}(\mathbf{X}) &= \sum_{n} \int\frac{\textrm{d}\varepsilon}{2\pi i}  \,f_n(\varepsilon) \textrm{tr}\Big\{ \Pi_n \left[ \partial_\alpha S^{\dagger}_{\mathbf{X}}(\varepsilon)  A^{\beta}_{\mathbf{X}}(\varepsilon) 
   -  A^{\beta \dagger}_{\mathbf{X}}(\varepsilon)    \partial_\alpha S_{\mathbf{X}}(\varepsilon) \right]    \Big\}_\mathrm{\rm{s}} \,.\label{eq:NEQ_friction}
 \end{align}
The white-noise fluctuating force correlator is given by
\begin{align}\label{eq:D_alphabeta}
 D_{\alpha\beta}(\mathbf{X}) =  \int\frac{\mathrm{d}\varepsilon}{2\pi}   \sum_{k,m} f_{m}(\varepsilon)\,\left[1\mp f_{k}(\varepsilon)\right] \, \textrm{tr}\Big\{ \Pi_m \Big[  S^{\dagger}_{\mathbf{X}}(\varepsilon) \partial_\alpha S_{\mathbf{X}}(\varepsilon) \Big]^\dagger  \cdot \Pi_k \cdot S^{\dagger}_{\mathbf{X}}(\varepsilon) \, \partial_\beta S_{\mathbf{X}}(\varepsilon)  \Big\} \,.
\end{align}
In equilibrium it is connected to the friction coefficient via $D_{\alpha\beta} = 2\,k_B \,T \, \gamma_{\alpha \beta}^{\rm{eq}}$, where $\gamma_{\alpha \beta}^{\rm{eq}}$ is evaluated in equilibrium. This agrees with the fluctuation-dissipation theoerem in Eq.~(\ref{eq:fluct-diss}) in the classical limit $\omega \ll T$.

\section{Identification of friction and noise within scattering theory}\label{app:scattering2}
Below we derive, within the framework of scattering theory, Eqs.~(\ref{eq:gamma0_relation}) and (\ref{eq:gamma_1eq}) - (\ref{eq:D_neq}) of the main text, which give the connection of the Loschmidt echo to fluctuations and dissipation. The friction tensor and the noise correlator are given in Eqs.~\eqref{eq:EQ_friction} - \eqref{eq:D_alphabeta}. In linear response these quantities read
\begin{align}\label{eq:gamma_0_EQ}
 \left( \boldsymbol{\gamma}^{\mathrm{\rm{eq}}}_0\right)_{\alpha\beta} &= \frac{1}{4\pi} \textrm{tr} \left\{ \partial_\alpha S_t^\dagger(\mu) \cdot  \partial_\beta S_t(\mu) \right\}_\mathrm{\rm{s}}    \\
 \left(\boldsymbol{\gamma}^{\mathrm{\rm{eq}}}_1\right)_{\alpha\beta} &= \frac{\Delta\mu}{8\pi} \left\{  \partial_\varepsilon  \left[ \left( \partial_\alpha S_t^\dagger(\varepsilon) \cdot  \partial_\beta S_t(\varepsilon)  \right)_{LL} - \left( \partial_\alpha S_t^\dagger(\varepsilon) \cdot  \partial_\beta S_t(\varepsilon)  \right)_{RR} \right]_{\varepsilon=\mu}  \right\}_\mathrm{\rm{s}} \label{eq:gamma_1_EQ} \\
 \left(\boldsymbol{\gamma}^{\mathrm{\rm{neq}}}_1 \right)_{\alpha\beta} &=  \frac{\Delta\mu}{4\pi i}   \Big\{  \left(  \partial_{\alpha} S_t^\dagger(\mu) \cdot A_t^{\beta}(\mu)   -  A_t^{\beta \dagger}(\mu) \cdot \partial_{\alpha} S_t(\mu)   \right)_{LL} -  \, \left(  \partial_{\alpha} S_t^\dagger(\mu) \cdot A_t^{\beta}(\mu)   -  A_t^{\beta \dagger}(\mu) \cdot \partial_{\alpha} S_t(\mu)   \right)_{RR} \Big\}_\mathrm{\rm{s}}  \label{eq:gamma_1_NEQ} \,.
\end{align}
  
We begin by expressing the function $K^{\alpha\beta}_{kn}(\varepsilon)$, defined in the main text in Eq.~(\ref{eq:def_F}), in terms of the S--matrix. By twice inserting the complete set of advanced scattering states $1 = \int\frac{\textrm{d}\varepsilon}{2\pi} \sum_m |\psi_m^{-}(\varepsilon) \rangle\langle \psi_m^{-}(\varepsilon) |$ we conclude from Eqs.~(\ref{eq:Ham_scat_states}) and~(\ref{eq:def_F}) that
\begin{align}\label{eq:K_def_app}
 K^{\alpha\beta}_{kn}(\varepsilon) &= \Big\{ \langle \psi_k^{\mathbf{X} +}(\varepsilon) | \partial_\alpha V_{\mathbf{X}} |  \psi_n^{\mathbf{X} +}(\varepsilon) \rangle \langle \psi_n^{\mathbf{X} +}(\varepsilon) | \partial_\beta V_{\mathbf{X}} |  \psi_k^{\mathbf{X} +}(\varepsilon) \rangle \Big\}_\mathrm{\rm{s}}  \vphantom{\int} \notag \\
 &= \left\{ \left( \partial_\alpha S_t^\dagger(\varepsilon) \cdot  S_t(\varepsilon) \right)_{kn}    \left( S_t^\dagger(\varepsilon) \cdot  \partial_\beta S_t(\varepsilon) \right)_{nk}  \right\}_\mathrm{\rm{s}} \,.
\end{align}
Because of the unitarity of the scattering matrix it readily follows that
\begin{align}
  \frac{1}{4\pi} \sum_{kn}  K^{\alpha\beta}_{kn}(\mu)     =  \frac{1}{4\pi} \textrm{tr} \left\{ \partial_\alpha S_t^\dagger(\mu) \cdot  \partial_\beta S_t(\mu) \right\}_\mathrm{\rm{s}}  =  \left( \boldsymbol{\gamma}_0^{\rm{eq}} \right)_{\alpha\beta}
\end{align}
at $T\rightarrow 0$ by referring to Eq.~(\ref{eq:gamma_0_EQ}). We thus have proven Eq.~(\ref{eq:gamma0_relation}). We continue with the derivation of Eq.~(\ref{eq:gamma_1eq}). Hereto we note that $\partial_\varepsilon^{\rm{s}} K^{\alpha\beta}_{kn}(\varepsilon) = \frac{1}{2}\,\partial_{\varepsilon}K^{\alpha\beta}_{kn}(\varepsilon)$ and $  \partial^{\rm{s}}_{\varepsilon} K^{\alpha\beta}_{kn}(\varepsilon)  =  \partial^{\rm{s}}_{\varepsilon} K^{\alpha\beta}_{nk}(\varepsilon) $. The latter property follows immediately from the relation $ K_{kn}^{\alpha\beta}(\varepsilon)  = K_{nk}^{\alpha\beta}(\varepsilon)$ and due to the symmetric summation in the indices $\alpha$ and $\beta$. Hence we get
\begin{align}
\frac{\Delta\mu}{4\pi}  \left[ \partial^{\rm{s}}_{\varepsilon} K^{\alpha\beta}_{LL}(\mu) -      \partial^{\rm{s}}_{\varepsilon} K^{\alpha\beta}_{RR}(\mu)\right]  &= \frac{\Delta\mu}{8\pi}   \sum_{n=L,R} \partial_\varepsilon \left[ K^{\alpha\beta}_{Ln}(\varepsilon) -  K^{\alpha\beta}_{Rn}(\varepsilon)  \right]_{\varepsilon=\mu}   \notag \\
&=\frac{\Delta\mu}{8\pi} \partial_\varepsilon  \left\{ \left[ \left( \partial_\alpha S_t^\dagger(\varepsilon) \cdot  \partial_\beta S_t(\varepsilon)  \right)_{LL} - \left( \partial_\alpha S_t^\dagger(\varepsilon)  \partial_\beta S_t(\varepsilon)  \right)_{RR} \right]_{\varepsilon=\mu} \right\}_\mathrm{\rm{s}}    \notag \\
&= \left(\boldsymbol{\gamma}_1^{\rm{eq}}\right)_{\alpha\beta}
\end{align}
by referring to Eq.~(\ref{eq:gamma_1_EQ}) in the last step which proves Eq.~(\ref{eq:gamma_1eq}).
In order to derive the relation in Eq.~(\ref{eq:gamma_1neq}) we first observe that $\partial^{\rm{a}}_\varepsilon K_{kn}^{\alpha\beta}(\varepsilon)$ can be written in terms of the S- and A-matrix. Again, inserting two complete sets of advanced scattering states, we find
\begin{align}
  \partial^{\rm{a}} K^{\alpha\beta}_{kn}(\varepsilon) &= \frac{1}{2} \left( \partial_{\varepsilon} - \partial_{\varepsilon'} \right)  K^{\alpha\beta}_{kn}(\varepsilon,\varepsilon')   \Big|_{\varepsilon'=\varepsilon}     \notag  \\
 &=\frac{1}{2}  \left\{ \langle \partial_{\varepsilon} \psi_k^{\mathbf{X} +}(\varepsilon) | \partial_\alpha V_{\mathbf{X}} |  \psi_n^{\mathbf{X} +}(\varepsilon) \rangle \langle \psi_n^{\mathbf{X} +}(\varepsilon) | \partial_\beta V_{\mathbf{X}} |  \psi_k^{\mathbf{X} +}(\varepsilon) \rangle  \right. \notag \\
 &\quad + \langle \psi_k^{\mathbf{X} +}(\varepsilon) | \partial_\alpha V_{\mathbf{X}} |  \psi_n^{\mathbf{X} +}(\varepsilon) \rangle \langle \psi_n^{\mathbf{X} +}(\varepsilon) | \partial_\beta V_{\mathbf{X}} | \partial_{\varepsilon} \psi_k^{\mathbf{X} +}(\varepsilon) \rangle  \vphantom{\int}  \notag \\
 &\quad - \left.\langle \psi_k^{\mathbf{X} +}(\varepsilon) | \partial_\alpha V_{\mathbf{X}} \, \partial_{\varepsilon} \Big[  |  \psi_n^{\mathbf{X} +}(\varepsilon) \rangle \langle \psi_n^{\mathbf{X} +}(\varepsilon) | \Big]  \, \partial_\beta V_{\mathbf{X}} |  \psi_k^{\mathbf{X} +}(\varepsilon) \rangle       \right\}_\mathrm{\rm{s}}  \notag  \\
  &=- \frac{1}{2}    \sum_{ml}      \left\{  \partial_\alpha S_t^{\dagger,km}(\varepsilon) \, \partial_\varepsilon \left[  S_t^{mn}(\varepsilon) \, S_t^{\dagger,nl}(\varepsilon) \right] \partial_\beta S_t^{lk}(\varepsilon)    \right\}_\mathrm{\rm{s}}       \notag \\
  &\quad - \frac{1}{i}    \sum_{ml}      \left\{   \left( \partial_\alpha S_t^\dagger(\varepsilon)    \cdot  S_t(\varepsilon) \right)_{kn}  \left( S_t^\dagger(\varepsilon)                A_t^{\beta}(\varepsilon)  \right)_{nk}  \right. \notag \\
  &\quad \left. -   \left( A_t^{\beta\,\dagger}(\varepsilon)   \cdot    S_t(\varepsilon) \right)_{kn} \left(S_t^\dagger(\varepsilon)  \cdot   \partial_\alpha S_t(\varepsilon)   \right)_{nk}  \right\}_\mathrm{\rm{s}}     \,.
\end{align}
In order to identify the pure nonequilibrium friction tensor in linear response [cf. Eq.~(\ref{eq:gamma_1_NEQ})], we observe that $   \partial^{\rm{a}}_{\varepsilon} K^{\alpha\beta}_{kn}(\varepsilon) = - \partial^{\rm{a}}_{\varepsilon} K^{\alpha\beta}_{nk}(\varepsilon)$, where we stress that $ \partial^{\rm{a}}_{\varepsilon} K^{\alpha\beta}_{nn}(\varepsilon)  = 0$. Restricting to two leads $k,n = L,R$ we thus conclude
\begin{align}
\frac{\Delta\mu}{4\pi}  \left[ \partial^{\rm{a}}_{\varepsilon} K^{\alpha\beta}_{RL}(\mu) -      \partial^{\rm{a}}_{\varepsilon} K^{\alpha\beta}_{LR}(\mu)\right]  &= \frac{\Delta\mu}{4\pi} \sum_{n=L,R}  \left[  \partial^{\rm{a}}_{\varepsilon} K^{\alpha\beta}_{Rn}(\mu) - \partial^{\rm{a}}_{\varepsilon} K^{\alpha\beta}_{Ln}(\mu)       \right]   \notag \\
 &=  \frac{\Delta\mu}{4\pi i}  \Big\{  \left(  \partial_{\alpha} S_t^\dagger(\mu) \cdot A_t^{\beta}(\mu)   -  A_t^{\beta \dagger}(\mu) \cdot \partial_{\alpha} S_t(\mu)   \right)_{LL} \notag  \\
 &\quad -  \, \left(  \partial_{\alpha} S_t^\dagger(\mu) \cdot A_t^{\beta}(\mu)   -  A_t^{\beta \dagger}(\mu) \cdot \partial_{\alpha} S_t(\mu)   \right)_{RR} \Big\}_\mathrm{\rm{s}} \cdot  \notag \\
 &= \left( \boldsymbol{\gamma}_1^{\rm{neq}} \right)_{\alpha\beta} \,.
\end{align}
Here we made use of $\textrm{tr}\{ \partial_{\alpha} S_t^\dagger   A_t^{\beta}   -  A_t^{\beta \dagger}   \partial_{\alpha} S_t  \} = 0$ to realize that the equilibrium contribution in $\gamma_{\alpha\beta}^{\rm{neq}}$ vanishes. We conclude with Eq.~(\ref{eq:gamma_1neq}). Finally, we show Eq.~(\ref{eq:D_neq}). For two leads, we write the correlator of the fluctuating force as~\cite{Bode:2012,Thomas:2012}
\begin{align}
D_{\alpha\beta}(\mathbf{X})  =  \int\frac{\mathrm{d}\varepsilon}{2\pi}   \sum_{k,m=L,R} f_{m}(\varepsilon)\,\left[1 - f_{k}(\varepsilon)\right]\, K^{\alpha\beta}_{mk}(\varepsilon) \,.
\end{align}
Making use of $f_k(\varepsilon) \,[1-f_k(\varepsilon)] = - T \, \partial_\varepsilon f_k(\varepsilon) = T\,\delta(\varepsilon-\mu_k)$ for small $T$, we can express the correlator in linear response to first order in the applied bias voltage as
\begin{align}
 D_{\alpha\beta}(\mathbf{X})   &=  \frac{T}{2\pi}    \left( \sum_{k,m=L,R}  K^{\alpha\beta}_{mk}(\mu)   +\frac{\Delta\mu}{2} \left[ \partial_\varepsilon K^{\alpha\beta}_{LL}(\varepsilon) - \partial_\varepsilon K^{\alpha\beta}_{RR}(\varepsilon) \right]_{\varepsilon=\mu} \right)   + \frac{\Delta\mu}{4\pi} \,  \left[  K^{\alpha\beta}_{LR}(\mu) + K^{\alpha\beta}_{RL}(\mu) \right]  
\end{align}
since $ K^{\alpha\beta}_{km}(\varepsilon)  = K^{\alpha\beta}_{mk}(\varepsilon)$. With $\frac{1}{2}\,\partial_\varepsilon = \partial^{\rm{s}}_{\varepsilon}$ we identify the equilibrium friction tensor in Eq.~(\ref{eq:gamma_0_EQ}) and its non-equilibrium correction in Eq.~(\ref{eq:gamma_1_EQ}) in the above expression for the correlator. For $T \rightarrow 0$ we then obtain Eq.~\eqref{eq:D_neq}.

\section{Alternative Keldysh approach}\label{app:Keldysh}

This appendix presents an alternative derivation of the results in the main text by using Keldysh Green's functions technique. We consider the generic Hamiltonian given in Eq.~\eqref{eq:Niels_Hamiltonian} and we define the \textit{time-dependent} Green functions of the dot
\begin{align}
 \mathcal{G}_{mm'}^R(t,t') &= - i \theta(t-t') \, \langle \{ d_m(t), d_{m'}^\dagger(t')    \} \rangle \\
 \mathcal{G}_{mm'}^A(t,t') &= i \theta(t'-t) \, \langle \{ d_m(t), d_{m'}^\dagger(t')    \} \rangle \\
 \mathcal{G}_{mm'}^>(t,t') &= - i  \, \langle  d_m(t) \, d_{m'}^\dagger(t')    \rangle \\
 \mathcal{G}_{mm'}^<(t,t') &=  i  \, \langle   d_{m'}^\dagger(t') \, d_m(t)    \rangle 
 \end{align}
where the curly bracktes~$\{\ldots,\ldots\}$ denote the anti-commutator operation. We assume stationary states throughout this section so that the above defined Green's functions depend on the difference of the time arguments. With these definitions the noise correlator defined in Eq.~(\ref{eq:ColorNoise}) reads~\cite{Bode:2011, Bode:2012}
\begin{align}
   D_{\alpha\beta}(t-t')  = \int \frac{\textrm{d}\varepsilon}{2\pi}\int \frac{\textrm{d}\varepsilon'}{2\pi} \, e^{i (\varepsilon-\varepsilon') (t-t')}  \textrm{tr}\Big\{ \Lambda_\alpha  \, \mathcal{G}^>(\varepsilon) \, \Lambda_\beta \, \mathcal{G}^<(\varepsilon') \Big\}_\mathrm{\rm{s}} \,.
\end{align}
with $\Lambda_\alpha = \partial_{X_\alpha} h_0(\mathbf{X})$. The functions $\mathcal{G}^>(\varepsilon)$ and $\mathcal{G}^<(\varepsilon')$ represent the Fourier transforms of the greater and lesser functions. It is sufficient to evaluate the noise correlator in the adiabatic limit as this already guarantees that the fluctuation-dissipation theorem is fulfilled. Hereto we introduce the Fourier transform of the adiabatic lesser and greater Green functions~$G^>(\varepsilon)$ and~$G^>(\varepsilon)$ with respect to a frozen configuration~$\mathbf{X}$ and conclude for the Fourier transform of the fluctuating force
\begin{align}
  D_{\alpha\beta}(\omega) = \int \frac{\textrm{d}\varepsilon}{2\pi} \, \textrm{tr}\Big\{ \Lambda_\alpha  \, G^>\Big(\varepsilon-\frac{\omega}{2}\Big) \, \Lambda_\beta \, G^<\Big(\varepsilon+\frac{\omega}{2}\Big) \Big\}_\mathrm{\rm{s}} \,. \label{eq:D_adiab_Green_fct}
\end{align}
Analogously we introduce the adiabatic retarded and advanced Green functions. Using Langreth rule, we can express the greater and the lesser Green function, respectively as
\begin{align}\label{eq:Langreth_rule_1}
 G^<(\varepsilon) &= G^R(\varepsilon) \, \Sigma^<(\varepsilon) \, G^A(\varepsilon)\\
 G^>(\varepsilon)&= G^R(\varepsilon) \, \Sigma^>(\varepsilon) \, G^A(\varepsilon) \label{eq:Langreth_rule_2}
\end{align}
with greater and lesser self-energies
\begin{align}
 \Sigma^<(\varepsilon) &=  \, i \, \sum_k f_k(\varepsilon) \, W^\dagger(\varepsilon) \, \Pi_k(\varepsilon) \, W(\varepsilon)\\
 \Sigma^>(\varepsilon) &= - i \, \sum_k (1- f_k(\varepsilon) )\, W^\dagger(\varepsilon) \, \Pi_k(\varepsilon) \, W(\varepsilon) 
\end{align}
where $\Pi_k(\varepsilon) = | \phi_k(\varepsilon) \rangle \langle \phi_k(\varepsilon) |$ is a projector onto lead space and the coupling matrix~$W$ is defined via the Hamiltonian in Eq.~(\ref{eq:Niels_Hamiltonian_4}).

Plugging Eqs.~(\ref{eq:Langreth_rule_1}) and~(\ref{eq:Langreth_rule_2}) into Eq.~(\ref{eq:D_adiab_Green_fct}) we can write the noise correlator as
\begin{align}\label{eq:D_K_tilde}
  D_{\alpha\beta}(\omega) &= \int \frac{\textrm{d}\varepsilon}{2\pi} \, \sum_{kn} f_k\Big(\varepsilon-\frac{\omega}{2}\Big) \,\left(  1 - f_n\Big(\varepsilon+\frac{\omega}{2}\Big) \right)  \tilde{K}^{\alpha\beta}_{kn}\Big(\varepsilon-\frac{\omega}{2},\varepsilon+\frac{\omega}{2}\Big)
\end{align}
where we defined the function $\tilde{K}_{kn}^{\alpha\beta}(\varepsilon,\varepsilon')$ as
\begin{align}\label{eq:def_K_Keldysh}
 \tilde{K}_{kn}^{\alpha\beta}(\varepsilon,\varepsilon') &= \textrm{tr} \left\{ \vphantom{\int} \Lambda_\alpha  \, G^R(\varepsilon)\, W^\dagger(\varepsilon) \, \Pi_k(\varepsilon) \, W(\varepsilon) \, G^A(\varepsilon)  \Lambda_\beta \, G^R(\varepsilon')\, W^\dagger(\varepsilon') \, \Pi_n(\varepsilon') \, W(\varepsilon') \, G^A(\varepsilon') \vphantom{\int} \right\}_\mathrm{\rm{s}} \,.
\end{align}
The expression in Eq.~(\ref{eq:D_K_tilde}) has the same structure as the expression in the main text in Eq.~(\ref{eq:D_omega_gen}). Here we proceed with the analogous steps to get to the result for the noise correlator in Eqs.~\eqref{eq:D_omega_sym1} and~\eqref{eq:D_omega_sym1} with all its consequences for the Loschmidt echo and the fidelity amplitude. In order to show their equivalence we are thus left with identifying the coefficients
\begin{align}\label{eq:coefficients_K_Keldysh_1}
\left( \boldsymbol{\gamma}_0^{\rm{eq}} \right)_{\alpha\beta} &=  \frac{1}{4\pi} \sum_{kn} \tilde{K}^{\alpha\beta}_{kn}(\mu)   \\
\left( \boldsymbol{\gamma}_1^{\rm{eq}} \right)_{\alpha\beta}  &=  \frac{\Delta\mu}{4\pi} \left( \partial^{\rm{s}}_{\varepsilon}\tilde{K}^{\alpha\beta}_{LL}(\mu) + \partial^{\rm{s}}_{\varepsilon}\tilde{K}^{\alpha\beta}_{RR}(\mu) \right)    \label{eq:coefficients_K_Keldysh_3}\\
\left( \boldsymbol{\gamma}_1^{\rm{neq}} \right)_{\alpha\beta} &=  \frac{\Delta\mu}{4\pi} \left( \partial^{\rm{a}}_{\varepsilon}\tilde{K}^{\alpha\beta}_{RL}(\mu) - \partial^{\rm{a}}_{\varepsilon}\tilde{K}^{\alpha\beta}_{LR}(\mu) \right)     \label{eq:coefficients_K_Keldysh_4} \\
\left( \frac{\mathbf{D}}{\Delta\mu} \right)_{\alpha\beta} &=   \frac{1}{4\pi} \left(\tilde{K}^{\alpha\beta}_{LR}(\mu) + \tilde{K}^{\alpha\beta}_{RL}(\mu)  \right)     \label{eq:coefficients_K_Keldysh_2}
\end{align}
for two leads under symmetric summation with respect to the indices~$\alpha$ and~$\beta$ in linear response at zero temperature [cf. ~Eqs.~(\ref{eq:gamma0_relation}), (\ref{eq:gamma_1eq}) - (\ref{eq:D_neq})]. We defined $\tilde{K}_{kn}^{\alpha\beta}(\varepsilon) = \tilde{K}_{kn}^{\alpha\beta}(\varepsilon,\varepsilon)$.

We begin with the identification of the friction tensor. We take the expression of the friction tensor in terms of the adiabatic Green functions from Refs.~\cite{Bode:2011, Bode:2012}
\begin{align}
 \gamma_{\alpha\beta} =  \int \frac{\textrm{d}\varepsilon}{2\pi} \, \textrm{tr}\Big\{ \Lambda_\alpha  \, G^>(\varepsilon) \, \Lambda_\beta \, \partial_\varepsilon G^<(\varepsilon) \Big\}_\mathrm{\rm{s}} \,.
\end{align}
We immediately conclude for the friction tensor in equilibrium that
\begin{align}
\left(\boldsymbol{\gamma}_0^{\rm{eq}}\right)_{\alpha\beta}&=\int \frac{\textrm{d}\varepsilon}{4\pi} \,\sum_{kn=L,R} \textrm{tr}\left\{ \vphantom{\sum} \Lambda_\alpha \,  G^R(\varepsilon) W^\dagger(\mu) \, \Pi_k(\mu) \, W(\mu)   \vphantom{\sum}  G^A(\mu)  \, \Lambda_\beta \,  G^R(\mu) \, W^\dagger(\mu) \, \Pi_n(\mu) \, W(\mu) \, G^A(\mu) \right\}_\mathrm{\rm{s}} 
\end{align}
since $f_k(\varepsilon) \, (1-f_k(\varepsilon)) = - T \, \partial_\varepsilon f_k(\varepsilon) = 0$ for $T \rightarrow 0$ as well as $f(\mu) = \frac{1}{2}$ and $-\partial_\varepsilon f(\varepsilon) = \delta(\varepsilon-\mu)$. A comparison to the definition in Eq.~(\ref{eq:def_K_Keldysh}) readily results in Eq.~(\ref{eq:coefficients_K_Keldysh_1}).

Next we address Eqs.~(\ref{eq:coefficients_K_Keldysh_3}) and~(\ref{eq:coefficients_K_Keldysh_4}). Hereto we write the Fermi functions of the left and right lead as $f_{L/R}(\varepsilon) = f(\varepsilon) \mp \frac{\Delta\mu}{2} \, \partial_\varepsilon f(\varepsilon) $. Keeping only terms linear in~$\Delta\mu$ we conclude after performing an integration by parts
\begin{align}
 (\boldsymbol{\gamma}_1^{\rm{eq}}  +\boldsymbol{\gamma}_1^{\rm{neq}})_{\alpha\beta}& =  \frac{\Delta\mu}{8\pi} \, \partial_\varepsilon \, \left(  \tilde{K}_{LL}^{\alpha\beta}(\mu,\varepsilon) + \tilde{K}_{LR}^{\alpha\beta}(\varepsilon,\mu)  \right)_{\varepsilon=\mu} - \frac{\Delta\mu}{8\pi} \, \partial_\varepsilon \, \left(  \tilde{K}_{RL}^{\alpha\beta}(\mu,\varepsilon) + \tilde{K}_{RR}^{\alpha\beta}(\varepsilon,\mu)  \right)_{\varepsilon=\mu}   \\\nonumber
& + \frac{\Delta\mu}{8\pi} \, \partial_\varepsilon \, \left(  \tilde{K}_{LL}^{\alpha\beta}(\mu,\varepsilon) + \tilde{K}_{RL}^{\alpha\beta}(\varepsilon,\mu)  \right)_{\varepsilon=\mu} - \frac{\Delta\mu}{8\pi} \, \partial_\varepsilon \, \left(  \tilde{K}_{LR}^{\alpha\beta}(\mu,\varepsilon) + \tilde{K}_{RR}^{\alpha\beta}(\varepsilon,\mu)  \right)_{\varepsilon=\mu}   \\\nonumber
&+ \int\frac{\textrm{d}\varepsilon}{2\pi}  \left[ \partial_\varepsilon f(\varepsilon) \right)^2 \,  \left( \tilde{K}^{\alpha\beta}_{LL}(\varepsilon) + \tilde{K}^{\alpha\beta}_{LR}(\varepsilon) -\tilde{K}^{\alpha\beta}_{RL}(\varepsilon) -\tilde{K}^{\alpha\beta}_{RR}(\varepsilon) -\tilde{K}^{\alpha\beta}_{LL}(\varepsilon) -\tilde{K}^{\alpha\beta}_{RL}(\varepsilon) +\tilde{K}^{\alpha\beta}_{LR}(\varepsilon) + \tilde{K}^{\alpha\beta}_{RR}(\varepsilon)  \right]  \,.
\end{align}
The last terms vanish due to the symmetric summation over $\alpha$ and $\beta$ and since $\tilde{K}^{\alpha\beta}_{kn}(\varepsilon) = \tilde{K}^{\alpha\beta}_{nk}(\varepsilon)$. This yields Eqs.~(\ref{eq:coefficients_K_Keldysh_3}) and~(\ref{eq:coefficients_K_Keldysh_4}).

Finally we identify the delta-correlated noise and prove Eq.~(\ref{eq:coefficients_K_Keldysh_2}) in linear response. We rely on the expression
\begin{align}
 D_{\alpha\beta}(\omega) = \int \frac{\textrm{d}\varepsilon}{2\pi} \, \textrm{tr}\{ \Lambda_\alpha  \, G^>(\varepsilon) \, \Lambda_\beta \, G^<(\varepsilon) \}_\mathrm{\rm{s}}
\end{align}
obtained in the literature~\cite{Bode:2011, Bode:2012} in terms of the Green functions of the dot. Using Eqs.~(\ref{eq:Langreth_rule_1}) and~(\ref{eq:Langreth_rule_2}) we immediately identify
\begin{align}
 D_{\alpha\beta}(\omega) = \int \frac{\textrm{d}\varepsilon}{2\pi} \,\sum_{kn=L,R} f_k(\varepsilon) \, (1-f_n(\varepsilon)) \, \tilde{K}_{LR}^{\alpha\beta}(\varepsilon) 
\end{align}
To linear response we find Eq.~(\ref{eq:coefficients_K_Keldysh_2}) by using $\tilde{K}^{\alpha\beta}_{kn}(\varepsilon) = \tilde{K}^{\alpha\beta}_{nk}(\varepsilon)$ and the above stated relations for the Fermi functions.

We end this appendix by showing the direct equivalence between the function $\tilde{K}_{kn}^{\alpha\beta}(\varepsilon,\varepsilon')$ defined in Eq.~(\ref{eq:def_K_Keldysh}) and $K_{kn}^{\alpha\beta}(\varepsilon,\varepsilon')$ defined in Eq.~(\ref{eq:def_F}), that is 
\begin{align}
   K_{kn}^{\alpha\beta}(\varepsilon,\varepsilon')  &= \Big\{ \langle \psi^{\mathbf{X}+}_k(\varepsilon) | \, \partial_\alpha H_{\mathbf{X}} \, | \psi^{\mathbf{X}+}_n(\varepsilon' \rangle \langle \psi^{\mathbf{X}+}_n(\varepsilon') | \,\partial_\beta H_{\mathbf{X}} \,| \psi^{\mathbf{X}+}_k(\varepsilon) \rangle  \Big\}_\mathrm{\rm{s}}
\end{align}
since $\partial_\alpha H_{\mathbf{X}} = \partial_\alpha V_{\mathbf{X}}$. We note that we can relate the Green function of the dot and the scattering states via the Lippmann-Schwinger equation~\cite{Thomas:2012}
\begin{align}\label{eq:Lipp_dot}
   \Pi_D\,| \psi_\eta^{{\bf X}_t+}(\varepsilon) \rangle = G^{R}(\varepsilon) \, W^\dagger \,|\phi_\eta(\varepsilon)\rangle \vphantom{\int}
\end{align}
where $\Pi_D$ denotes a projector onto the space of the dot. With the aid of $\partial_\alpha H_{\mathbf{X}} = \Pi_D \, \partial_\alpha h_0(\mathbf{X}) \, \Pi_D$ and the Lippmann-Schwinger equation projected onto the dot's space in Eq.~(\ref{eq:Lipp_dot}) we have
\begin{align}
  K_{kn}^{\alpha\beta}(\varepsilon,\varepsilon')  &=  \Big\{ \langle \phi_k(\varepsilon) |\, W(\varepsilon) \, G^A(\varepsilon) \, \partial_\alpha H_{\mathbf{X}} \, G^R(\varepsilon') W^\dagger(\varepsilon') \, |\phi_n(\varepsilon') \rangle \langle \phi_n(\varepsilon') |  \, W(\varepsilon') \, G^A(\varepsilon')  \partial_\beta H_{\mathbf{X}} \, G^R(\varepsilon)\, W^\dagger(\varepsilon) \, |\phi_k(\varepsilon) \rangle \Big\}_\mathrm{\rm{s}} \,.
\end{align}
Due to symmetric summation under exchanging the indices $\alpha$ and $\beta$ and using that $\Pi_n(\varepsilon') = |\phi_n(\varepsilon') \rangle \langle \phi_n(\varepsilon') |$ we conclude that
\begin{align}
 K_{kn}^{\alpha\beta} (\varepsilon,\varepsilon')  &= \textrm{tr} \Big\{  \partial_\alpha H_{\mathbf{X}} \, G^R(\varepsilon') \, W^\dagger(\varepsilon') \, \Pi_n(\varepsilon') \, W(\varepsilon')     G^A(\varepsilon') \, \partial_\beta H_{\mathbf{X}} \,G^R(\varepsilon) \, W^\dagger(\varepsilon) \, \Pi_k(\varepsilon) \, W(\varepsilon) \, G^A(\varepsilon) \Big\}_\mathrm{\rm{s}} \notag \\
  &=  \tilde{K}_{kn}^{\alpha\beta}(\varepsilon,\varepsilon')  \,.
\end{align}
This shows explicitly the equivalence to the scattering states approach.
\end{widetext}
\end{appendices}

\bibliography{Loschmidt_BIB}

%merlin.mbs apsrev4-1.bst 2010-07-25 4.21a (PWD, AO, DPC) hacked
%Control: key (0)
%Control: author (8) initials jnrlst
%Control: editor formatted (1) identically to author
%Control: production of article title (-1) disabled
%Control: page (0) single
%Control: year (1) truncated
%Control: production of eprint (0) enabled
\begin{thebibliography}{52}%
\makeatletter
\providecommand \@ifxundefined [1]{%
 \@ifx{#1\undefined}
}%
\providecommand \@ifnum [1]{%
 \ifnum #1\expandafter \@firstoftwo
 \else \expandafter \@secondoftwo
 \fi
}%
\providecommand \@ifx [1]{%
 \ifx #1\expandafter \@firstoftwo
 \else \expandafter \@secondoftwo
 \fi
}%
\providecommand \natexlab [1]{#1}%
\providecommand \enquote  [1]{``#1''}%
\providecommand \bibnamefont  [1]{#1}%
\providecommand \bibfnamefont [1]{#1}%
\providecommand \citenamefont [1]{#1}%
\providecommand \href@noop [0]{\@secondoftwo}%
\providecommand \href [0]{\begingroup \@sanitize@url \@href}%
\providecommand \@href[1]{\@@startlink{#1}\@@href}%
\providecommand \@@href[1]{\endgroup#1\@@endlink}%
\providecommand \@sanitize@url [0]{\catcode `\\12\catcode `\$12\catcode
  `\&12\catcode `\#12\catcode `\^12\catcode `\_12\catcode `\%12\relax}%
\providecommand \@@startlink[1]{}%
\providecommand \@@endlink[0]{}%
\providecommand \url  [0]{\begingroup\@sanitize@url \@url }%
\providecommand \@url [1]{\endgroup\@href {#1}{\urlprefix }}%
\providecommand \urlprefix  [0]{URL }%
\providecommand \Eprint [0]{\href }%
\providecommand \doibase [0]{http://dx.doi.org/}%
\providecommand \selectlanguage [0]{\@gobble}%
\providecommand \bibinfo  [0]{\@secondoftwo}%
\providecommand \bibfield  [0]{\@secondoftwo}%
\providecommand \translation [1]{[#1]}%
\providecommand \BibitemOpen [0]{}%
\providecommand \bibitemStop [0]{}%
\providecommand \bibitemNoStop [0]{.\EOS\space}%
\providecommand \EOS [0]{\spacefactor3000\relax}%
\providecommand \BibitemShut  [1]{\csname bibitem#1\endcsname}%
\let\auto@bib@innerbib\@empty
%</preamble>
\bibitem [{\citenamefont {Caldeira}\ and\ \citenamefont
  {Leggett}(1981)}]{Caldeira:1981}%
  \BibitemOpen
  \bibfield  {author} {\bibinfo {author} {\bibfnamefont {A.~O.}\ \bibnamefont
  {Caldeira}}\ and\ \bibinfo {author} {\bibfnamefont {A.~J.}\ \bibnamefont
  {Leggett}},\ }\href {\doibase 10.1103/PhysRevLett.46.211} {\bibfield
  {journal} {\bibinfo  {journal} {Phys. Rev. Lett.}\ }\textbf {\bibinfo
  {volume} {46}},\ \bibinfo {pages} {211} (\bibinfo {year} {1981})}\BibitemShut
  {NoStop}%
\bibitem [{\citenamefont {Caldeira}\ and\ \citenamefont
  {Leggett}(1983)}]{Caldeira:1983}%
  \BibitemOpen
  \bibfield  {author} {\bibinfo {author} {\bibfnamefont {A.~O.}\ \bibnamefont
  {Caldeira}}\ and\ \bibinfo {author} {\bibfnamefont {A.~J.}\ \bibnamefont
  {Leggett}},\ }\href@noop {} {\bibfield  {journal} {\bibinfo  {journal}
  {Annals of Physics}\ }\textbf {\bibinfo {volume} {149}},\ \bibinfo {pages}
  {374} (\bibinfo {year} {1983})}\BibitemShut {NoStop}%
\bibitem [{\citenamefont {Devoret}\ and\ \citenamefont
  {Schoelkopf}(2013)}]{DevoretScience13}%
  \BibitemOpen
  \bibfield  {author} {\bibinfo {author} {\bibfnamefont {M.~H.}\ \bibnamefont
  {Devoret}}\ and\ \bibinfo {author} {\bibfnamefont {R.~J.}\ \bibnamefont
  {Schoelkopf}},\ }\href {\doibase 10.1126/science.1231930} {\bibfield
  {journal} {\bibinfo  {journal} {Science}\ }\textbf {\bibinfo {volume}
  {339}},\ \bibinfo {pages} {1169} (\bibinfo {year} {2013})}\BibitemShut
  {NoStop}%
\bibitem [{\citenamefont {Awschalom}\ \emph {et~al.}(2013)\citenamefont
  {Awschalom}, \citenamefont {Bassett}, \citenamefont {Dzurak}, \citenamefont
  {Hu},\ and\ \citenamefont {Petta}}]{AwschalomScience13}%
  \BibitemOpen
  \bibfield  {author} {\bibinfo {author} {\bibfnamefont {D.~D.}\ \bibnamefont
  {Awschalom}}, \bibinfo {author} {\bibfnamefont {L.~C.}\ \bibnamefont
  {Bassett}}, \bibinfo {author} {\bibfnamefont {A.~S.}\ \bibnamefont {Dzurak}},
  \bibinfo {author} {\bibfnamefont {E.~L.}\ \bibnamefont {Hu}}, \ and\ \bibinfo
  {author} {\bibfnamefont {J.~R.}\ \bibnamefont {Petta}},\ }\href {\doibase
  10.1126/science.1231364} {\bibfield  {journal} {\bibinfo  {journal}
  {Science}\ }\textbf {\bibinfo {volume} {339}},\ \bibinfo {pages} {1174}
  (\bibinfo {year} {2013})}\BibitemShut {NoStop}%
\bibitem [{\citenamefont {Weiss}(1999)}]{Weiss:1999}%
  \BibitemOpen
  \bibfield  {author} {\bibinfo {author} {\bibfnamefont {U.}~\bibnamefont
  {Weiss}},\ }\href@noop {} {\emph {\bibinfo {title} {Quantum Dissipative
  Systems}}},\ Vol.~\bibinfo {volume} {10}\ (\bibinfo  {publisher} {World
  Scientific Publishing Co. Pte. Ltd.},\ \bibinfo {address} {Singapore},\
  \bibinfo {year} {1999})\BibitemShut {NoStop}%
\bibitem [{\citenamefont {Peres}(1984)}]{Peres:1984}%
  \BibitemOpen
  \bibfield  {author} {\bibinfo {author} {\bibfnamefont {A.}~\bibnamefont
  {Peres}},\ }\href {\doibase 10.1103/PhysRevA.30.1610} {\bibfield  {journal}
  {\bibinfo  {journal} {Phys. Rev. A}\ }\textbf {\bibinfo {volume} {30}},\
  \bibinfo {pages} {1610} (\bibinfo {year} {1984})}\BibitemShut {NoStop}%
\bibitem [{\citenamefont {Gorin}\ \emph {et~al.}(2006)\citenamefont {Gorin},
  \citenamefont {Prosen}, \citenamefont {Seligman},\ and\ \citenamefont
  {\ifmmode \check{Z}\else \v{Z}\fi{}nidari\ifmmode~\check{c}\else
  \v{c}\fi{}}}]{Gorin:2006_PhysRep}%
  \BibitemOpen
  \bibfield  {author} {\bibinfo {author} {\bibfnamefont {T.}~\bibnamefont
  {Gorin}}, \bibinfo {author} {\bibfnamefont {T.}~\bibnamefont {Prosen}},
  \bibinfo {author} {\bibfnamefont {T.~H.}\ \bibnamefont {Seligman}}, \ and\
  \bibinfo {author} {\bibfnamefont {M.}~\bibnamefont {\ifmmode \check{Z}\else
  \v{Z}\fi{}nidari\ifmmode~\check{c}\else \v{c}\fi{}}},\ }\href {\doibase
  10.1016/j.physrep.2006.09.003} {\bibfield  {journal} {\bibinfo  {journal}
  {Phys. Rep.}\ }\textbf {\bibinfo {volume} {435}},\ \bibinfo {pages} {33}
  (\bibinfo {year} {2006})},\ \bibinfo {note} {and references
  therein}\BibitemShut {NoStop}%
\bibitem [{\citenamefont {Lesovik}\ \emph {et~al.}(2006)\citenamefont
  {Lesovik}, \citenamefont {Hassler},\ and\ \citenamefont
  {Blatter}}]{Lesovik:2006}%
  \BibitemOpen
  \bibfield  {author} {\bibinfo {author} {\bibfnamefont {G.~B.}\ \bibnamefont
  {Lesovik}}, \bibinfo {author} {\bibfnamefont {F.}~\bibnamefont {Hassler}}, \
  and\ \bibinfo {author} {\bibfnamefont {G.}~\bibnamefont {Blatter}},\ }\href
  {\doibase 10.1103/PhysRevLett.96.106801} {\bibfield  {journal} {\bibinfo
  {journal} {Phys. Rev. Lett.}\ }\textbf {\bibinfo {volume} {96}},\ \bibinfo
  {pages} {106801} (\bibinfo {year} {2006})}\BibitemShut {NoStop}%
\bibitem [{\citenamefont {Karkuszewski}\ \emph {et~al.}(2002)\citenamefont
  {Karkuszewski}, \citenamefont {Jarzynski},\ and\ \citenamefont
  {Zurek}}]{Karkuszewski:2002}%
  \BibitemOpen
  \bibfield  {author} {\bibinfo {author} {\bibfnamefont {Z.~P.}\ \bibnamefont
  {Karkuszewski}}, \bibinfo {author} {\bibfnamefont {C.}~\bibnamefont
  {Jarzynski}}, \ and\ \bibinfo {author} {\bibfnamefont {W.~H.}\ \bibnamefont
  {Zurek}},\ }\href {\doibase 10.1103/PhysRevLett.89.170405} {\bibfield
  {journal} {\bibinfo  {journal} {Phys. Rev. Lett.}\ }\textbf {\bibinfo
  {volume} {89}},\ \bibinfo {pages} {170405} (\bibinfo {year}
  {2002})}\BibitemShut {NoStop}%
\bibitem [{\citenamefont {Goold}\ \emph {et~al.}(2011)\citenamefont {Goold},
  \citenamefont {Fogarty}, \citenamefont {Lo~Gullo}, \citenamefont
  {Paternostro},\ and\ \citenamefont {Busch}}]{Goold:2011}%
  \BibitemOpen
  \bibfield  {author} {\bibinfo {author} {\bibfnamefont {J.}~\bibnamefont
  {Goold}}, \bibinfo {author} {\bibfnamefont {T.}~\bibnamefont {Fogarty}},
  \bibinfo {author} {\bibfnamefont {N.}~\bibnamefont {Lo~Gullo}}, \bibinfo
  {author} {\bibfnamefont {M.}~\bibnamefont {Paternostro}}, \ and\ \bibinfo
  {author} {\bibfnamefont {T.}~\bibnamefont {Busch}},\ }\href {\doibase
  10.1103/PhysRevA.84.063632} {\bibfield  {journal} {\bibinfo  {journal} {Phys.
  Rev. A}\ }\textbf {\bibinfo {volume} {84}},\ \bibinfo {pages} {063632}
  (\bibinfo {year} {2011})}\BibitemShut {NoStop}%
\bibitem [{\citenamefont {Anderson}(1967)}]{Anderson:1967}%
  \BibitemOpen
  \bibfield  {author} {\bibinfo {author} {\bibfnamefont {P.~W.}\ \bibnamefont
  {Anderson}},\ }\href {\doibase 10.1103/PhysRevLett.18.1049} {\bibfield
  {journal} {\bibinfo  {journal} {Phys. Rev. Lett.}\ }\textbf {\bibinfo
  {volume} {18}},\ \bibinfo {pages} {1049} (\bibinfo {year}
  {1967})}\BibitemShut {NoStop}%
\bibitem [{\citenamefont {Mahan}(1967)}]{Mahan:1967}%
  \BibitemOpen
  \bibfield  {author} {\bibinfo {author} {\bibfnamefont {G.~D.}\ \bibnamefont
  {Mahan}},\ }\href {\doibase 10.1103/PhysRev.163.612} {\bibfield  {journal}
  {\bibinfo  {journal} {Phys. Rev.}\ }\textbf {\bibinfo {volume} {163}},\
  \bibinfo {pages} {612} (\bibinfo {year} {1967})}\BibitemShut {NoStop}%
\bibitem [{\citenamefont {Nozi\`eres}\ and\ \citenamefont
  {de~Dominicis}(1969)}]{Dominicis:1969}%
  \BibitemOpen
  \bibfield  {author} {\bibinfo {author} {\bibfnamefont {P.}~\bibnamefont
  {Nozi\`eres}}\ and\ \bibinfo {author} {\bibfnamefont {C.~T.}\ \bibnamefont
  {de~Dominicis}},\ }\href {\doibase 10.1103/PhysRev.178.1097} {\bibfield
  {journal} {\bibinfo  {journal} {Phys. Rev.}\ }\textbf {\bibinfo {volume}
  {178}},\ \bibinfo {pages} {1097} (\bibinfo {year} {1969})}\BibitemShut
  {NoStop}%
\bibitem [{\citenamefont {Rivier}\ and\ \citenamefont
  {Simanek}(1971)}]{Rivier:1971}%
  \BibitemOpen
  \bibfield  {author} {\bibinfo {author} {\bibfnamefont {N.}~\bibnamefont
  {Rivier}}\ and\ \bibinfo {author} {\bibfnamefont {E.}~\bibnamefont
  {Simanek}},\ }\href {\doibase 10.1103/PhysRevLett.26.435} {\bibfield
  {journal} {\bibinfo  {journal} {Phys. Rev. Lett.}\ }\textbf {\bibinfo
  {volume} {26}},\ \bibinfo {pages} {435} (\bibinfo {year} {1971})}\BibitemShut
  {NoStop}%
\bibitem [{\citenamefont {Ng}(1995)}]{Ng:1995}%
  \BibitemOpen
  \bibfield  {author} {\bibinfo {author} {\bibfnamefont {T.-K.}\ \bibnamefont
  {Ng}},\ }\href {\doibase 10.1103/PhysRevB.51.2009} {\bibfield  {journal}
  {\bibinfo  {journal} {Phys. Rev. B}\ }\textbf {\bibinfo {volume} {51}},\
  \bibinfo {pages} {2009} (\bibinfo {year} {1995})}\BibitemShut {NoStop}%
\bibitem [{\citenamefont {Ng}(1996)}]{Ng:1996}%
  \BibitemOpen
  \bibfield  {author} {\bibinfo {author} {\bibfnamefont {T.-K.}\ \bibnamefont
  {Ng}},\ }\href {\doibase 10.1103/PhysRevB.54.5814} {\bibfield  {journal}
  {\bibinfo  {journal} {Phys. Rev. B}\ }\textbf {\bibinfo {volume} {54}},\
  \bibinfo {pages} {5814} (\bibinfo {year} {1996})}\BibitemShut {NoStop}%
\bibitem [{\citenamefont {Abanin}\ and\ \citenamefont
  {Levitov}(2004)}]{Levitov:2004}%
  \BibitemOpen
  \bibfield  {author} {\bibinfo {author} {\bibfnamefont {D.~A.}\ \bibnamefont
  {Abanin}}\ and\ \bibinfo {author} {\bibfnamefont {L.~S.}\ \bibnamefont
  {Levitov}},\ }\href {\doibase 10.1103/PhysRevLett.93.126802} {\bibfield
  {journal} {\bibinfo  {journal} {Phys. Rev. Lett.}\ }\textbf {\bibinfo
  {volume} {93}},\ \bibinfo {pages} {126802} (\bibinfo {year}
  {2004})}\BibitemShut {NoStop}%
\bibitem [{\citenamefont {Abanin}\ and\ \citenamefont
  {Levitov}(2005)}]{Levitov:2005}%
  \BibitemOpen
  \bibfield  {author} {\bibinfo {author} {\bibfnamefont {D.~A.}\ \bibnamefont
  {Abanin}}\ and\ \bibinfo {author} {\bibfnamefont {L.~S.}\ \bibnamefont
  {Levitov}},\ }\href {\doibase 10.1103/PhysRevLett.94.186803} {\bibfield
  {journal} {\bibinfo  {journal} {Phys. Rev. Lett.}\ }\textbf {\bibinfo
  {volume} {94}},\ \bibinfo {pages} {186803} (\bibinfo {year}
  {2005})}\BibitemShut {NoStop}%
\bibitem [{\citenamefont {Sols}\ and\ \citenamefont
  {Guinea}(1987)}]{Sols:1987}%
  \BibitemOpen
  \bibfield  {author} {\bibinfo {author} {\bibfnamefont {F.}~\bibnamefont
  {Sols}}\ and\ \bibinfo {author} {\bibfnamefont {F.}~\bibnamefont {Guinea}},\
  }\href {\doibase 10.1103/PhysRevB.36.7775} {\bibfield  {journal} {\bibinfo
  {journal} {Phys. Rev. B}\ }\textbf {\bibinfo {volume} {36}},\ \bibinfo
  {pages} {7775} (\bibinfo {year} {1987})}\BibitemShut {NoStop}%
\bibitem [{Note1()}]{Note1}%
  \BibitemOpen
  \bibinfo {note} {The small-distance limit corresponds to small variations in
  the classical coordinates.}\BibitemShut {Stop}%
\bibitem [{\citenamefont {Sch\"onhammer}(1991)}]{Schoenhammer:1991}%
  \BibitemOpen
  \bibfield  {author} {\bibinfo {author} {\bibfnamefont {K.}~\bibnamefont
  {Sch\"onhammer}},\ }\href {\doibase 10.1103/PhysRevB.43.11323} {\bibfield
  {journal} {\bibinfo  {journal} {Phys. Rev. B}\ }\textbf {\bibinfo {volume}
  {43}},\ \bibinfo {pages} {11323} (\bibinfo {year} {1991})}\BibitemShut
  {NoStop}%
\bibitem [{\citenamefont {Muzykantskii}\ \emph
  {et~al.}(2003{\natexlab{a}})\citenamefont {Muzykantskii}, \citenamefont
  {d'Ambrumenil},\ and\ \citenamefont {Braunecker}}]{Braunecker:2003}%
  \BibitemOpen
  \bibfield  {author} {\bibinfo {author} {\bibfnamefont {B.}~\bibnamefont
  {Muzykantskii}}, \bibinfo {author} {\bibfnamefont {N.}~\bibnamefont
  {d'Ambrumenil}}, \ and\ \bibinfo {author} {\bibfnamefont {B.}~\bibnamefont
  {Braunecker}},\ }\href {\doibase 10.1103/PhysRevLett.91.266602} {\bibfield
  {journal} {\bibinfo  {journal} {Phys. Rev. Lett.}\ }\textbf {\bibinfo
  {volume} {91}},\ \bibinfo {pages} {266602} (\bibinfo {year}
  {2003}{\natexlab{a}})}\BibitemShut {NoStop}%
\bibitem [{\citenamefont {Segal}\ \emph {et~al.}(2007)\citenamefont {Segal},
  \citenamefont {Reichman},\ and\ \citenamefont {Millis}}]{Segal:2007}%
  \BibitemOpen
  \bibfield  {author} {\bibinfo {author} {\bibfnamefont {D.}~\bibnamefont
  {Segal}}, \bibinfo {author} {\bibfnamefont {D.~R.}\ \bibnamefont {Reichman}},
  \ and\ \bibinfo {author} {\bibfnamefont {A.~J.}\ \bibnamefont {Millis}},\
  }\href {\doibase 10.1103/PhysRevB.76.195316} {\bibfield  {journal} {\bibinfo
  {journal} {Phys. Rev. B}\ }\textbf {\bibinfo {volume} {76}},\ \bibinfo
  {pages} {195316} (\bibinfo {year} {2007})}\BibitemShut {NoStop}%
\bibitem [{\citenamefont {Bode}\ \emph {et~al.}(2011)\citenamefont {Bode},
  \citenamefont {Kusminskiy}, \citenamefont {Egger},\ and\ \citenamefont {von
  Oppen}}]{Bode:2011}%
  \BibitemOpen
  \bibfield  {author} {\bibinfo {author} {\bibfnamefont {N.}~\bibnamefont
  {Bode}}, \bibinfo {author} {\bibfnamefont {S.~V.}\ \bibnamefont
  {Kusminskiy}}, \bibinfo {author} {\bibfnamefont {R.}~\bibnamefont {Egger}}, \
  and\ \bibinfo {author} {\bibfnamefont {F.}~\bibnamefont {von Oppen}},\ }\href
  {\doibase 10.1103/PhysRevLett.107.036804} {\bibfield  {journal} {\bibinfo
  {journal} {Phys. Rev. Lett.}\ }\textbf {\bibinfo {volume} {107}},\ \bibinfo
  {pages} {036804} (\bibinfo {year} {2011})}\BibitemShut {NoStop}%
\bibitem [{\citenamefont {Thomas}\ \emph {et~al.}(2012)\citenamefont {Thomas},
  \citenamefont {Karzig}, \citenamefont {Kusminskiy}, \citenamefont
  {Zar\'and},\ and\ \citenamefont {von Oppen}}]{Thomas:2012}%
  \BibitemOpen
  \bibfield  {author} {\bibinfo {author} {\bibfnamefont {M.}~\bibnamefont
  {Thomas}}, \bibinfo {author} {\bibfnamefont {T.}~\bibnamefont {Karzig}},
  \bibinfo {author} {\bibfnamefont {S.~V.}\ \bibnamefont {Kusminskiy}},
  \bibinfo {author} {\bibfnamefont {G.}~\bibnamefont {Zar\'and}}, \ and\
  \bibinfo {author} {\bibfnamefont {F.}~\bibnamefont {von Oppen}},\ }\href
  {\doibase 10.1103/PhysRevB.86.195419} {\bibfield  {journal} {\bibinfo
  {journal} {Phys. Rev. B}\ }\textbf {\bibinfo {volume} {86}},\ \bibinfo
  {pages} {195419} (\bibinfo {year} {2012})}\BibitemShut {NoStop}%
\bibitem [{\citenamefont {Berry}\ and\ \citenamefont
  {Robbins}(1993)}]{Berry:1993}%
  \BibitemOpen
  \bibfield  {author} {\bibinfo {author} {\bibfnamefont {M.~V.}\ \bibnamefont
  {Berry}}\ and\ \bibinfo {author} {\bibfnamefont {J.~M.}\ \bibnamefont
  {Robbins}},\ }\href {\doibase 10.1098/rspa.1993.0127} {\bibfield  {journal}
  {\bibinfo  {journal} {Proc. R. Soc. Lond. A}\ }\textbf {\bibinfo {volume}
  {442}},\ \bibinfo {pages} {659} (\bibinfo {year} {1993})}\BibitemShut
  {NoStop}%
\bibitem [{\citenamefont {Dundas}\ \emph {et~al.}(2009)\citenamefont {Dundas},
  \citenamefont {Mceniry},\ and\ \citenamefont {Todorov}}]{Todorov:2009}%
  \BibitemOpen
  \bibfield  {author} {\bibinfo {author} {\bibfnamefont {D.}~\bibnamefont
  {Dundas}}, \bibinfo {author} {\bibfnamefont {E.}~\bibnamefont {Mceniry}}, \
  and\ \bibinfo {author} {\bibfnamefont {T.~N.}\ \bibnamefont {Todorov}},\
  }\href {\doibase 10.1038/NNANO.2008.411} {\bibfield  {journal} {\bibinfo
  {journal} {Nature Nanotechnology}\ }\textbf {\bibinfo {volume} {4}},\
  \bibinfo {pages} {99} (\bibinfo {year} {2009})}\BibitemShut {NoStop}%
\bibitem [{\citenamefont {L\"u}\ \emph {et~al.}(2010)\citenamefont {L\"u},
  \citenamefont {Brandbyge},\ and\ \citenamefont
  {Hedeg\aa{a}rd}}]{Hedegard:2010}%
  \BibitemOpen
  \bibfield  {author} {\bibinfo {author} {\bibfnamefont {J.-T.}\ \bibnamefont
  {L\"u}}, \bibinfo {author} {\bibfnamefont {M.}~\bibnamefont {Brandbyge}}, \
  and\ \bibinfo {author} {\bibfnamefont {P.}~\bibnamefont {Hedeg\aa{a}rd}},\
  }\href {\doibase 10.1021/nl904233u} {\bibfield  {journal} {\bibinfo
  {journal} {Nano Lett.}\ }\textbf {\bibinfo {volume} {10}},\ \bibinfo {pages}
  {1657} (\bibinfo {year} {2010})}\BibitemShut {NoStop}%
\bibitem [{\citenamefont {Bode}\ \emph {et~al.}(2012)\citenamefont {Bode},
  \citenamefont {{Viola Kusminskiy}}, \citenamefont {Egger},\ and\
  \citenamefont {von Oppen}}]{Bode:2012}%
  \BibitemOpen
  \bibfield  {author} {\bibinfo {author} {\bibfnamefont {N.}~\bibnamefont
  {Bode}}, \bibinfo {author} {\bibfnamefont {S.}~\bibnamefont {{Viola
  Kusminskiy}}}, \bibinfo {author} {\bibfnamefont {R.}~\bibnamefont {Egger}}, \
  and\ \bibinfo {author} {\bibfnamefont {F.}~\bibnamefont {von Oppen}},\
  }\href@noop {} {\bibfield  {journal} {\bibinfo  {journal} {Beilstein J.
  Nanotechnol.}\ }\textbf {\bibinfo {volume} {3}},\ \bibinfo {pages} {144}
  (\bibinfo {year} {2012})}\BibitemShut {NoStop}%
\bibitem [{Note2()}]{Note2}%
  \BibitemOpen
  \bibinfo {note} {We note that we assume the initial Hamiltonian $\protect
  \mathcal {H}_{\protect \mathbf {X}} = \protect \mathcal {H}_0 + \protect
  \mathcal {V}_{\protect \mathbf {X}}$ to be time independent. The Green
  functions depend therefore only on the difference of the time arguments.
  Throughout the paper we define the Fourier transform of a function $f(t)$ as
  $f(\omega ) = \DOTSI \intop \ilimits@ _{-\infty }^{\infty } \protect \textrm
  {d}t \protect \tmspace +\thinmuskip {.1667em} e^{i \omega t} \protect
  \tmspace +\thinmuskip {.1667em} f(t) $ with its inverse $f(t) = \DOTSI \intop
  \ilimits@ _{-\infty }^{\infty } \protect \frac {\protect \textrm {d}t}{2\pi }
  \protect \tmspace +\thinmuskip {.1667em} e^{-i \omega t} \protect \tmspace
  +\thinmuskip {.1667em} f(\omega ) $.}\BibitemShut {Stop}%
\bibitem [{\citenamefont {Kubo}(1966)}]{Kubo:1966}%
  \BibitemOpen
  \bibfield  {author} {\bibinfo {author} {\bibfnamefont {R.}~\bibnamefont
  {Kubo}},\ }\href {\doibase 10.1088/0034-4885/29/1/306} {\bibfield  {journal}
  {\bibinfo  {journal} {Rep. Prog. Phys.}\ }\textbf {\bibinfo {volume} {29}},\
  \bibinfo {pages} {255} (\bibinfo {year} {1966})}\BibitemShut {NoStop}%
\bibitem [{Note3()}]{Note3}%
  \BibitemOpen
  \bibinfo {note} {Note that the quantities $\protect \mathbf {D}$ and
  $\protect \mathbf {\gamma }$ depend on $\protect \mathbf {X}$. In order to
  simplify the notation, we do not write this dependence
  explicitly.}\BibitemShut {Stop}%
\bibitem [{\citenamefont {Schotte}\ and\ \citenamefont
  {Schotte}(1969)}]{Schotte:1969}%
  \BibitemOpen
  \bibfield  {author} {\bibinfo {author} {\bibfnamefont {K.~D.}\ \bibnamefont
  {Schotte}}\ and\ \bibinfo {author} {\bibfnamefont {U.}~\bibnamefont
  {Schotte}},\ }\href {\doibase 10.1103/PhysRev.182.479} {\bibfield  {journal}
  {\bibinfo  {journal} {Phys. Rev.}\ }\textbf {\bibinfo {volume} {182}},\
  \bibinfo {pages} {479} (\bibinfo {year} {1969})}\BibitemShut {NoStop}%
\bibitem [{\citenamefont {Ogawa}\ \emph {et~al.}(1992)\citenamefont {Ogawa},
  \citenamefont {Furusaki},\ and\ \citenamefont {Nagaosa}}]{Nagaosa:1992}%
  \BibitemOpen
  \bibfield  {author} {\bibinfo {author} {\bibfnamefont {T.}~\bibnamefont
  {Ogawa}}, \bibinfo {author} {\bibfnamefont {A.}~\bibnamefont {Furusaki}}, \
  and\ \bibinfo {author} {\bibfnamefont {N.}~\bibnamefont {Nagaosa}},\ }\href
  {\doibase 10.1103/PhysRevLett.68.3638} {\bibfield  {journal} {\bibinfo
  {journal} {Phys. Rev. Lett.}\ }\textbf {\bibinfo {volume} {68}},\ \bibinfo
  {pages} {3638} (\bibinfo {year} {1992})}\BibitemShut {NoStop}%
\bibitem [{\citenamefont {Kane}\ \emph {et~al.}(1994)\citenamefont {Kane},
  \citenamefont {Matveev},\ and\ \citenamefont {Glazman}}]{Glazman:1994}%
  \BibitemOpen
  \bibfield  {author} {\bibinfo {author} {\bibfnamefont {C.~L.}\ \bibnamefont
  {Kane}}, \bibinfo {author} {\bibfnamefont {K.~A.}\ \bibnamefont {Matveev}}, \
  and\ \bibinfo {author} {\bibfnamefont {L.~I.}\ \bibnamefont {Glazman}},\
  }\href {\doibase 10.1103/PhysRevB.49.2253} {\bibfield  {journal} {\bibinfo
  {journal} {Phys. Rev. B}\ }\textbf {\bibinfo {volume} {49}},\ \bibinfo
  {pages} {2253} (\bibinfo {year} {1994})}\BibitemShut {NoStop}%
\bibitem [{\citenamefont {Imambekov}\ \emph {et~al.}(2012)\citenamefont
  {Imambekov}, \citenamefont {Schmidt},\ and\ \citenamefont
  {Glazman}}]{Imambekov:2012}%
  \BibitemOpen
  \bibfield  {author} {\bibinfo {author} {\bibfnamefont {A.}~\bibnamefont
  {Imambekov}}, \bibinfo {author} {\bibfnamefont {T.~L.}\ \bibnamefont
  {Schmidt}}, \ and\ \bibinfo {author} {\bibfnamefont {L.~I.}\ \bibnamefont
  {Glazman}},\ }\href {\doibase 10.1103/RevModPhys.84.1253} {\bibfield
  {journal} {\bibinfo  {journal} {Rev. Mod. Phys.}\ }\textbf {\bibinfo {volume}
  {84}},\ \bibinfo {pages} {1253} (\bibinfo {year} {2012})}\BibitemShut
  {NoStop}%
\bibitem [{\citenamefont {Yuval}\ and\ \citenamefont
  {Anderson}(1970)}]{Yuval:1970}%
  \BibitemOpen
  \bibfield  {author} {\bibinfo {author} {\bibfnamefont {G.}~\bibnamefont
  {Yuval}}\ and\ \bibinfo {author} {\bibfnamefont {P.~W.}\ \bibnamefont
  {Anderson}},\ }\href {\doibase 10.1103/PhysRevB.1.1522} {\bibfield  {journal}
  {\bibinfo  {journal} {Phys. Rev. B}\ }\textbf {\bibinfo {volume} {1}},\
  \bibinfo {pages} {1522} (\bibinfo {year} {1970})}\BibitemShut {NoStop}%
\bibitem [{\citenamefont {Knap}\ \emph {et~al.}(2012)\citenamefont {Knap},
  \citenamefont {Shashi}, \citenamefont {Nishida}, \citenamefont {Imambekov},
  \citenamefont {Abanin},\ and\ \citenamefont {Demler}}]{Knap:2012}%
  \BibitemOpen
  \bibfield  {author} {\bibinfo {author} {\bibfnamefont {M.}~\bibnamefont
  {Knap}}, \bibinfo {author} {\bibfnamefont {A.}~\bibnamefont {Shashi}},
  \bibinfo {author} {\bibfnamefont {Y.}~\bibnamefont {Nishida}}, \bibinfo
  {author} {\bibfnamefont {A.}~\bibnamefont {Imambekov}}, \bibinfo {author}
  {\bibfnamefont {D.~A.}\ \bibnamefont {Abanin}}, \ and\ \bibinfo {author}
  {\bibfnamefont {E.}~\bibnamefont {Demler}},\ }\href {\doibase
  10.1103/PhysRevX.2.041020} {\bibfield  {journal} {\bibinfo  {journal} {Phys.
  Rev. X}\ }\textbf {\bibinfo {volume} {2}},\ \bibinfo {pages} {041020}
  (\bibinfo {year} {2012})}\BibitemShut {NoStop}%
\bibitem [{Note4()}]{Note4}%
  \BibitemOpen
  \bibinfo {note} {In a scattering approach, ``small changes in the potential''
  can be understood as $\tau _D\delta \protect \mathcal {H}_{\protect \bf X}\ll
  1$, since the electronic Hamiltonian is of the order of $1/\tau _D$. For the
  adiabatic quench the change in the potential is time dependent such that most
  of the scattering electrons see a change smaller than $\delta \protect
  \mathcal {H}_{\protect \bf X}$, hence the condition is also
  satisfied.}\BibitemShut {Stop}%
\bibitem [{Note5()}]{Note5}%
  \BibitemOpen
  \bibinfo {note} {Note that $\protect \mathbf {D}(\omega )$ is not the Fourier
  transform of $\protect \mathbf {D}(t-t')$. It is rather the symmetrized
  Fourier transform (in frequency~$\omega $).}\BibitemShut {Stop}%
\bibitem [{\citenamefont {M\"under}\ \emph {et~al.}(2012)\citenamefont
  {M\"under}, \citenamefont {Weichselbaum}, \citenamefont {Goldstein},
  \citenamefont {Gefen},\ and\ \citenamefont {von Delft}}]{Muender:2012}%
  \BibitemOpen
  \bibfield  {author} {\bibinfo {author} {\bibfnamefont {W.}~\bibnamefont
  {M\"under}}, \bibinfo {author} {\bibfnamefont {A.}~\bibnamefont
  {Weichselbaum}}, \bibinfo {author} {\bibfnamefont {M.}~\bibnamefont
  {Goldstein}}, \bibinfo {author} {\bibfnamefont {Y.}~\bibnamefont {Gefen}}, \
  and\ \bibinfo {author} {\bibfnamefont {J.}~\bibnamefont {von Delft}},\ }\href
  {\doibase 10.1103/PhysRevB.85.235104} {\bibfield  {journal} {\bibinfo
  {journal} {Phys. Rev. B}\ }\textbf {\bibinfo {volume} {85}},\ \bibinfo
  {pages} {235104} (\bibinfo {year} {2012})}\BibitemShut {NoStop}%
\bibitem [{Note6()}]{Note6}%
  \BibitemOpen
  \bibinfo {note} {Note that these relations are valid for $\tau /\tau _D \gg
  1$.}\BibitemShut {Stop}%
\bibitem [{\citenamefont {D\'ora}\ \emph {et~al.}(2013)\citenamefont {D\'ora},
  \citenamefont {Pollmann}, \citenamefont {Fort\'agh},\ and\ \citenamefont
  {Zar\'and}}]{Zarand:2013}%
  \BibitemOpen
  \bibfield  {author} {\bibinfo {author} {\bibfnamefont {B.}~\bibnamefont
  {D\'ora}}, \bibinfo {author} {\bibfnamefont {F.}~\bibnamefont {Pollmann}},
  \bibinfo {author} {\bibfnamefont {J.}~\bibnamefont {Fort\'agh}}, \ and\
  \bibinfo {author} {\bibfnamefont {G.}~\bibnamefont {Zar\'and}},\ }\href
  {\doibase 10.1103/PhysRevLett.111.046402} {\bibfield  {journal} {\bibinfo
  {journal} {Phys. Rev. Lett.}\ }\textbf {\bibinfo {volume} {111}},\ \bibinfo
  {pages} {046402} (\bibinfo {year} {2013})}\BibitemShut {NoStop}%
\bibitem [{\citenamefont {Sachdeva}\ \emph {et~al.}(2013)\citenamefont
  {Sachdeva}, \citenamefont {Nag}, \citenamefont {Agarwal},\ and\ \citenamefont
  {Dutta}}]{Sachdeva:2013}%
  \BibitemOpen
  \bibfield  {author} {\bibinfo {author} {\bibfnamefont {R.}~\bibnamefont
  {Sachdeva}}, \bibinfo {author} {\bibfnamefont {T.}~\bibnamefont {Nag}},
  \bibinfo {author} {\bibfnamefont {A.}~\bibnamefont {Agarwal}}, \ and\
  \bibinfo {author} {\bibfnamefont {A.}~\bibnamefont {Dutta}},\ }\href@noop {}
  {\bibfield  {journal} {\bibinfo  {journal} {arXiv:1311.1926}\ } (\bibinfo
  {year} {2013})}\BibitemShut {NoStop}%
\bibitem [{\citenamefont {Aleiner}\ \emph {et~al.}(1997)\citenamefont
  {Aleiner}, \citenamefont {Wingreen},\ and\ \citenamefont
  {Meir}}]{Aleiner:1997}%
  \BibitemOpen
  \bibfield  {author} {\bibinfo {author} {\bibfnamefont {I.}~\bibnamefont
  {Aleiner}}, \bibinfo {author} {\bibfnamefont {N.}~\bibnamefont {Wingreen}}, \
  and\ \bibinfo {author} {\bibfnamefont {Y.}~\bibnamefont {Meir}},\ }\href
  {\doibase 10.1103/PhysRevLett.79.3740} {\bibfield  {journal} {\bibinfo
  {journal} {Phys. Rev. Lett.}\ }\textbf {\bibinfo {volume} {79}},\ \bibinfo
  {pages} {3740} (\bibinfo {year} {1997})}\BibitemShut {NoStop}%
\bibitem [{\citenamefont {L\"u}\ \emph {et~al.}(2012)\citenamefont {L\"u},
  \citenamefont {Brandbyge}, \citenamefont {Hedeg\aa{}rd}, \citenamefont
  {Todorov},\ and\ \citenamefont {Dundas}}]{Lue:2012}%
  \BibitemOpen
  \bibfield  {author} {\bibinfo {author} {\bibfnamefont {J.-T.}\ \bibnamefont
  {L\"u}}, \bibinfo {author} {\bibfnamefont {M.}~\bibnamefont {Brandbyge}},
  \bibinfo {author} {\bibfnamefont {P.}~\bibnamefont {Hedeg\aa{}rd}}, \bibinfo
  {author} {\bibfnamefont {T.~N.}\ \bibnamefont {Todorov}}, \ and\ \bibinfo
  {author} {\bibfnamefont {D.}~\bibnamefont {Dundas}},\ }\href {\doibase
  10.1103/PhysRevB.85.245444} {\bibfield  {journal} {\bibinfo  {journal} {Phys.
  Rev. B}\ }\textbf {\bibinfo {volume} {85}},\ \bibinfo {pages} {245444}
  (\bibinfo {year} {2012})}\BibitemShut {NoStop}%
\bibitem [{\citenamefont {B\"uttiker}(1992)}]{Buttiker:1992}%
  \BibitemOpen
  \bibfield  {author} {\bibinfo {author} {\bibfnamefont {M.}~\bibnamefont
  {B\"uttiker}},\ }\href {\doibase 10.1103/PhysRevB.46.12485} {\bibfield
  {journal} {\bibinfo  {journal} {Phys. Rev. B}\ }\textbf {\bibinfo {volume}
  {46}},\ \bibinfo {pages} {12485} (\bibinfo {year} {1992})}\BibitemShut
  {NoStop}%
\bibitem [{\citenamefont {Levinson}(2000)}]{Levinson:2000}%
  \BibitemOpen
  \bibfield  {author} {\bibinfo {author} {\bibfnamefont {Y.}~\bibnamefont
  {Levinson}},\ }\href {\doibase 10.1103/PhysRevB.61.4748} {\bibfield
  {journal} {\bibinfo  {journal} {Phys. Rev. B}\ }\textbf {\bibinfo {volume}
  {61}},\ \bibinfo {pages} {4748} (\bibinfo {year} {2000})}\BibitemShut
  {NoStop}%
\bibitem [{Note7()}]{Note7}%
  \BibitemOpen
  \bibinfo {note} {Note that the linear response expansion is indeed an
  expansion in $\Delta \mu \protect \tmspace +\thinmuskip {.1667em}\tau _D$
  since the energy derivatives appearing in Eqs.~\protect \textup {\hbox
  {\mathsurround \z@ \protect \normalfont (\ignorespaces \ref
  {eq:gamma_1eq}\unskip \@@italiccorr )}} and~\protect \textup {\hbox
  {\mathsurround \z@ \protect \normalfont (\ignorespaces \ref
  {eq:gamma_1neq}\unskip \@@italiccorr )}} are of order the Wigner time delay,
  which can be estimated to be of the order of $\tau _D$.}\BibitemShut {Stop}%
\bibitem [{\citenamefont {Muzykantskii}\ \emph
  {et~al.}(2003{\natexlab{b}})\citenamefont {Muzykantskii}, \citenamefont
  {d'Ambrumenil},\ and\ \citenamefont {Braunecker}}]{Muzykantskii:2003}%
  \BibitemOpen
  \bibfield  {author} {\bibinfo {author} {\bibfnamefont {B.}~\bibnamefont
  {Muzykantskii}}, \bibinfo {author} {\bibfnamefont {N.}~\bibnamefont
  {d'Ambrumenil}}, \ and\ \bibinfo {author} {\bibfnamefont {B.}~\bibnamefont
  {Braunecker}},\ }\href {\doibase 10.1103/PhysRevLett.91.266602} {\bibfield
  {journal} {\bibinfo  {journal} {Phys. Rev. Lett.}\ }\textbf {\bibinfo
  {volume} {91}},\ \bibinfo {pages} {266602} (\bibinfo {year}
  {2003}{\natexlab{b}})}\BibitemShut {NoStop}%
\bibitem [{\citenamefont {Moskalets}\ and\ \citenamefont
  {B\"uttiker}(2004)}]{Moskalets:2004}%
  \BibitemOpen
  \bibfield  {author} {\bibinfo {author} {\bibfnamefont {M.}~\bibnamefont
  {Moskalets}}\ and\ \bibinfo {author} {\bibfnamefont {M.}~\bibnamefont
  {B\"uttiker}},\ }\href {\doibase 10.1103/PhysRevB.69.205316} {\bibfield
  {journal} {\bibinfo  {journal} {Phys. Rev. B}\ }\textbf {\bibinfo {volume}
  {69}},\ \bibinfo {pages} {205316} (\bibinfo {year} {2004})}\BibitemShut
  {NoStop}%
\bibitem [{\citenamefont {Arrachea}\ and\ \citenamefont
  {Moskalets}(2006)}]{Arrachea:2006}%
  \BibitemOpen
  \bibfield  {author} {\bibinfo {author} {\bibfnamefont {L.}~\bibnamefont
  {Arrachea}}\ and\ \bibinfo {author} {\bibfnamefont {M.}~\bibnamefont
  {Moskalets}},\ }\href {\doibase 10.1103/PhysRevB.74.245322} {\bibfield
  {journal} {\bibinfo  {journal} {Phys. Rev. B}\ }\textbf {\bibinfo {volume}
  {74}},\ \bibinfo {pages} {245322} (\bibinfo {year} {2006})}\BibitemShut
  {NoStop}%
\end{thebibliography}%

\end{document}